\pdfoutput=1
\documentclass[aps,prb,reprint,superscriptaddress,nofootinbib]{revtex4-1}

\usepackage{graphicx}
\usepackage{amsmath}
\usepackage{amssymb}
\usepackage{amsfonts}
\usepackage{dcolumn}
\usepackage{dsfont}
\usepackage{latexsym}
\usepackage{rotating}
\usepackage{color}
\usepackage{latexsym}
\usepackage{bbm}
\usepackage{subfigure}
\usepackage{float}
\usepackage{epsfig}
\usepackage{psfrag}
\usepackage{natbib, hyperref}
\usepackage{bm}
\usepackage{amsthm}
\usepackage{eucal}
\usepackage{mathrsfs}
\usepackage{url}
\usepackage{braket}
\usepackage{ulem}
\usepackage{calligra}
\usepackage[T1]{fontenc}
\usepackage[utf8]{inputenc}
\usepackage{newunicodechar}
\usepackage{marvosym}
\usepackage[absolute]{textpos}
\usepackage[cal=cm]{mathalfa}
\usepackage{soul}

\usepackage{color}

\usepackage{hyperref}
\hypersetup{
colorlinks=true,final=true,
        linkcolor=blue,
        citecolor=blue,
        filecolor=blue,
        urlcolor=blue,
}

\usepackage{blindtext}
\makeatletter
\newcommand\emailx[1]{%
\move@AF%
\def\@affil{{\normalfont\,#1\strut}{}}%
}%
\makeatother

\begin{document}
\setlength{\abovedisplayskip}{3pt}
\setlength{\belowdisplayskip}{3pt}

\setcounter{figure}{0}
\renewcommand{\figurename}{{\bf Fig.}}
\renewcommand{\thefigure}{{\bf \arabic{figure}}}

\title{Majorana corner states on the dice lattice}

\author{Narayan Mohanta}
\affiliation{Materials Science and Technology Division, Oak Ridge National Laboratory, Oak Ridge, TN 37831, USA}
\affiliation{Department of Physics and Astronomy, The University of Tennessee, Knoxville, TN 37996, USA}
\affiliation{Department of Physics, Indian Institute of Technology Roorkee, Roorkee 247667, India}

\author{Rahul Soni}
\affiliation{Materials Science and Technology Division, Oak Ridge National Laboratory, Oak Ridge, TN 37831, USA}
\affiliation{Department of Physics and Astronomy, The University of Tennessee, Knoxville, TN 37996, USA}

\author{Satoshi Okamoto}
\affiliation{Materials Science and Technology Division, Oak Ridge National Laboratory, Oak Ridge, TN 37831, USA}

\author{Elbio Dagotto}
\affiliation{Materials Science and Technology Division, Oak Ridge National Laboratory, Oak Ridge, TN 37831, USA}
\affiliation{Department of Physics and Astronomy, The University of Tennessee, Knoxville, TN 37996, USA}

\begin{abstract}
Lattice geometry continues providing exotic topological phases in condensed matter physics. Exciting recent examples are the higher-order topological phases, manifesting via localized lower-dimensional boundary states. Moreover, flat electronic bands with a non-trivial topology arise in various lattices and can hold a finite superfluid density, bounded by the Chern number $C$. Here we consider attractive interaction in the dice lattice that hosts flat bands with $C\!=\!\pm2$ and show that the induced superconducting state exhibits a second-order topological phase with mixed singlet-triplet pairing. The second-order nature of the topological superconducting phase is revealed by the zero-energy Majorana bound states at the lattice corners. Hence, the topology of the normal state dictates the nature of the Majorana localization. These findings suggest that flat bands with a higher Chern number provide feasible platforms for inducing higher-order topological superconductivity.
\end{abstract}

\maketitle

\section*{Introduction}
\vspace{-1em}
Higher order topology in quantum matter has recently generated a flurry of activity in several broad areas, including the field of superconductivity~\cite{Schindler_SciAdv2018,PhysRevLett.119.246401,PhysRevLett.120.026801,PhysRevB.97.205136,PhysRevB.98.201114,PhysRevB.97.205135,PhysRevX.9.011012,Li_2DMater2021,Ghosh_PRB2021,PhysRevResearch.2.012060,PhysRevResearch.2.043300}. At the boundaries and vortex cores, 
a topological superconductor harbors Majorana quasiparticles, with potential value in the long-sought area of decoherence-free quantum computing~\cite{Kitaev_2001,RevModPhys.80.1083,Sarma2015,PhysRevX.6.031016,RevModPhys.87.137,Mohanta_EPL2014}. Topological superconductivity can be induced, for example, by a Rashba spin-orbit coupling together with a magnetic field~\cite{PhysRevLett.105.077001,Mourik_Science2012,Rokhinson2012,Deng_Science2016}, and also by a spatially-modulated spin texture in proximity to a conventional superconductor~\cite{Desjardins2019,PhysRevApplied.12.034048,Mohanta_CommPhys2021,Herbrych2021}. A $n^{\rm th}$-order topological superconductor in $d$ dimensions hosts $(d\!-\!n)$-dimensional Majorana states~\cite{Song_PRL2017}. The corner-localized Majorana bound states (MBS) in a second-order two-dimensional topological superconductor are particularly interesting because a two-dimensional array of corner MBS -- useful for demonstrating non-Abelian statistics -- can be easily achieved~\cite{PhysRevB.98.165144,PhysRevResearch.2.032068,PhysRevResearch.2.043025}. These corner MBS have been proposed in many platforms including a topological insulator in proximity to a $d$-wave or $s^{\pm}$-wave superconductor, extended Hubbard model with spin-orbit coupling~\cite{Kheirkhah_PRL2020} and a Josephson junction bilayer~\cite{PhysRevLett.121.096803,PhysRevLett.121.186801,PhysRevLett.122.126402}.

Following the discovery of unconventional superconductivity in twisted-bilayer graphene~\cite{Cao2018}, a series of studies suggested the possibility of electronic pairing from repulsive interaction in materials with a high density of states at the Fermi level, such as in a flat electronic band, leading to superconductivity with a high critical temperature~\cite{PhysRevB.101.014501,PhysRevLett.126.027002,Volovik2016,Aoki2020,PhysRevB.106.125155}. When a flat band is topologically nontrivial, the topological invariant places a lower bound on the superfluid weight $D_s$ \textit{i.e.} $D_s \! \ge \!C$, where $C$ is the Chern number of the flat band~\cite{Peotta2015,PhysRevLett.124.167002}. In this case, near a band-inversion wavevector, the Berry phase can convert a repulsive interaction between two oppositely-moving electrons into an effective attraction. Therefore, the connection between the topology of the normal state and the induced superconductivity has remained as an important subject of investigation, especially in the presence of a repulsive interaction~\cite{Sticlet_PRB2014,Verma_PNAS2021}.

\begin{figure*}[t]
\begin{center}
\vspace{0mm}
\epsfig{file=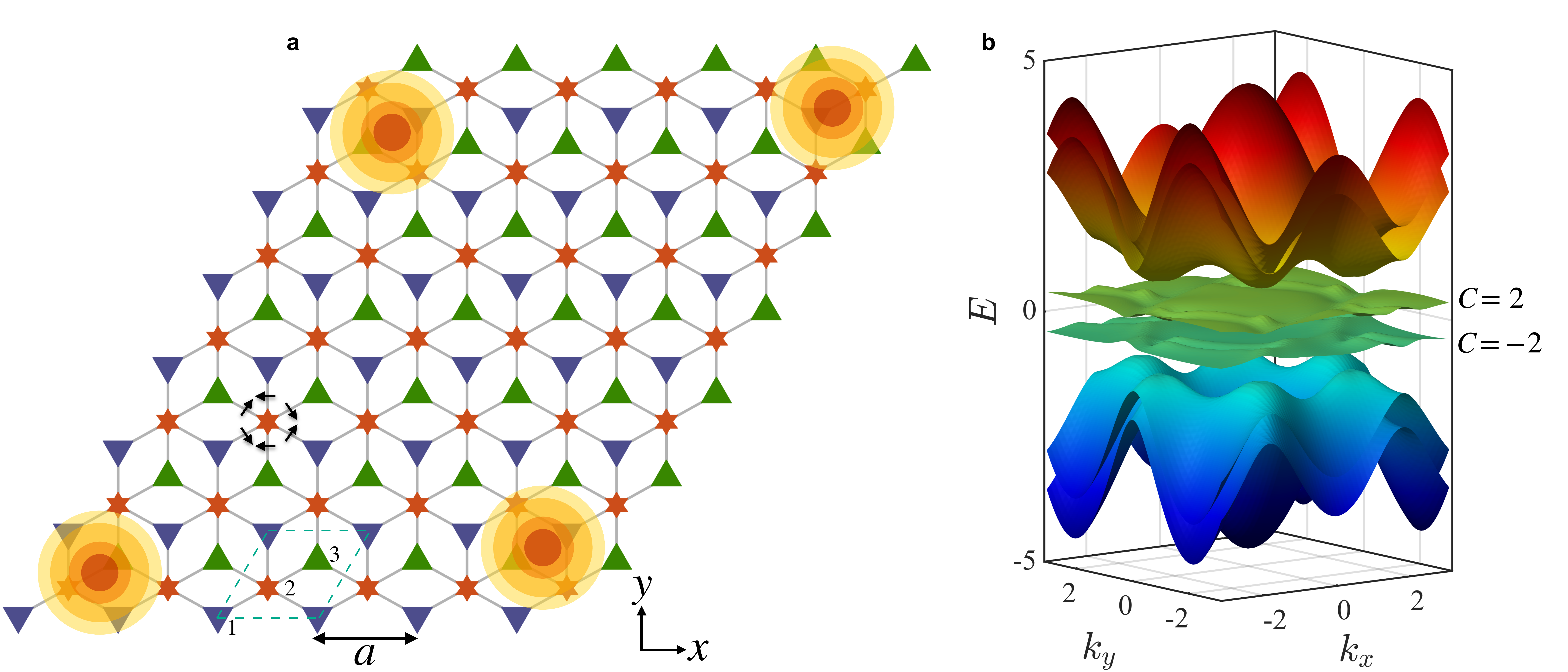,trim=0.0in 0.0in 0.0in 0.0in,clip=false, width=160mm}
\caption{{\bf Majorana corner states and topological flat bands in the dice lattice.} {\bf {\small a}} Schematic description of the corner-localized Majorana bound states on the Dice lattice with open boundaries in the second-order topological superconducting phase. There are three inequivalent sites in the unit cell, shown by the dashed lines. The triangles denote three-coordination sites and the hexagrams denote six-coordination sites. The black arrows surrounding the six-coordination site (middle layer) represent the vectors of the Rashba spin-orbit coupling, with clockwise sense of rotation for the upper triangles (bottom  layer) and counter-clockwise for the lower triangles (top layer). {\bf {\small b}} Electronic bands of the dice lattice in the presence of a spin-orbit coupling and a magnetic field, showing the nearly-flat topological bands with Chern number $C\!=\!\pm2$, close to the Fermi level. The topological superconducting phase is obtained by populating the lower topological flat band at the Fermi level.}
\label{dice_flatbands}
\vspace{-0mm}
\end{center}
\end{figure*}

It is, however, mostly unclear whether the induced superconductivity in the topological flat bands is also topologically non-trivial. Here we consider the topological flat bands with $C\!=\!\pm2$ on the dice lattice~\cite{PhysRevB.84.241103,PhysRevB.102.045105} in the presence of an attractive interaction and show that second-order topological superconductivity is induced by populating a topological flat band at the Fermi level. A hallmark of the induced second-order topological superconducting phase is found via the zero-energy MBS, protected by mirror symmetry and localized at the lattice corners. 

The dice lattice and the four corner MBS are shown schematically in Fig.~\ref{dice_flatbands}{\bf {\small a}}. The bipartite nature of the dice lattice, which can be envisaged as two merged triangular lattices, protects two degenerate flat bands coexisting with four other dispersive bands. Such a geometry can be realized using a few layers of transition-metal oxides, dichalcogenides, and graphene. In the simplest realization of the dice lattice involving three (111) layers of cubic transition-metal oxides, such as in a SrTiO$_3$/SrIrO$_3$/SrTiO$_3$ trilayer, the cubic symmetry is reduced to trigonal symmetry. The strong spin-orbit coupling from the Ir$^{4+}$ ion and the broken inversion symmetry produces a Rashba spin-orbit coupling. In the reduced ${\cal D}$$_{3d}$ symmetry of the trilayer, the Rashba spin-orbit coupling vectors lie in the plane parallel to the trilayer and have opposite senses of rotation for the top and the bottom layers of the three-coordination sites, surrounding the middle layer of six-coordination sites, as shown in Fig.~\ref{dice_flatbands}{\bf {\small a}} by the black arrows. In the presence of this Rashba spin-orbit coupling, the flat bands become isolated from the dispersive bands. Repulsive interactions in the flat bands then spontaneously generate ferro/ferri-magnetic order on the Kramer's pair of flat bands~\cite{PhysRevB.84.241103,PhysRevB.102.045105}, especially when they are close to half filling, and split them into two nearly-flat bands with Chern number $C\!=\! \pm2$, as shown in Fig.~\ref{dice_flatbands}{\bf {\small b}}. A local four-fermion interaction, leading to an excitonic gap, may also generate two well-separated topological flat bands~\cite{Gorbar_PRB2021}.

Besides the topological origin of superconductivity in the flat bands, superconductivity in the transition-metal oxide trilayer can also be induced by doping SrTiO$_3$~\cite{PhysRevLett.12.474,PhysRev.163.380}, for example, by Nb. Our calculations reveal that the superconducting state realized in the  dice-lattice topological flat bands, exhibits both singlet and triplet pairings. Such a singlet-triplet mixing is allowed by broken inversion symmetry in this oxide trilayer. Using symmetry analysis, we find that the possible nearest-neihgbor pairing channels allowed in this lattice geometry are $d_{xy}$, $d_{x^2-y^2}$, $p_x$ and $p_y$. The nearest-neighbor pairing amplitudes were found to be complex numbers, indicating the chiral nature of the induced topological superconducting phase. Besides finding corner MBS, supporting the second-order nature of the topological superconducting phase, we perform an analysis of the quasiparticle excitation gap in momentum space and identify the parameter regime where the excitation gap becomes finite, that characterizes the induced topological superconducting phase in the topologically-non-trivial flat bands.

The dice lattice has been studied for decades for its intriguing electronic properties~\cite{Horiguchi_1974,Sutherland_PRB1986,Vidal_PRL1998,Vidal_PRB2001}.
It is a special case of the $\alpha \!-\! \tau_{_3}$ lattice which interpolates between the dice lattice ($\alpha \!=\!0$, pseudospin 1) and the honeycomb lattice  ($\alpha \!=\!1$, pseudospin 1/2)~\cite{PhysRevB.94.125435}. By changing the hopping parameter $\alpha$, the orbital susceptibility can be changed continuously from dia to paramagnetic~\cite{Raoux_PRL2014}. It is also possible to transform the honeycomb lattice into the dice lattice, and vice versa, in an experimentally-simulated ultracold atomic gas platform~\cite{PhysRevB.73.144511}. Also, the results presented here are relevant to possible topological superconducting phases in twisted bilayer/multilayer graphene~\cite{PhysRevLett.121.087001}.

\vspace{-1em}
\section*{Results}
\vspace{-1em}
\noindent  \textbf{\small Model and set up}\\
\indent The electron pairing in the topological flat bands of the dice lattice, realizable in a transition-metal oxide trilayer as discussed above, can be described by the following tight-binding Hamiltonian
\begin{align}
{\cal H}\!=\!&-t\! \sum_{\langle i\alpha,j\beta \rangle,\sigma} \!(c_{i\alpha\sigma}^{\dagger}c_{j\beta\sigma}^{\phantom{\dagger}}+{\rm H.c.})
\!-\mu\! \sum_{i,\alpha,\sigma} \! c_{i\alpha\sigma}^{\dagger}c_{i\alpha\sigma}^{\phantom{\dagger}} \nonumber \\
&-\lambda \! \sum_{\langle i\alpha,j\beta \rangle,\sigma,\sigma^{\prime}} \!(i [\mathbf{\hat{D}}_{ij}\cdot \boldsymbol{\sigma}]_{_{\alpha \beta}}^{\sigma \sigma^{\prime}}c_{i\alpha\sigma}^{\dagger}c_{j\beta\sigma^{\prime}}^{\phantom{\dagger}}+{\rm H.c.})  \nonumber \\
&-B_z \! \sum_{i,\alpha,\sigma,\sigma^{\prime}} \!( [\sigma_z]^{\sigma \sigma^{\prime}}c_{i\alpha\sigma}^{\dagger}c^{\phantom{\dagger}}_{i\alpha\sigma^{\prime}}+{\rm H.c.}) \nonumber \\
&-U_0 \sum_{i,\alpha} \!n_{i\alpha\uparrow}n_{i\alpha\downarrow}-\frac{U_1}{2}\sum_{i,j,\alpha,\beta,\sigma,\sigma^{\prime}} n_{i\alpha\sigma}n_{j\beta\sigma^{\prime}},
\label{Ham}
\end{align}
\begin{figure*}[t]
\begin{center}
\vspace{-0mm}
\epsfig{file=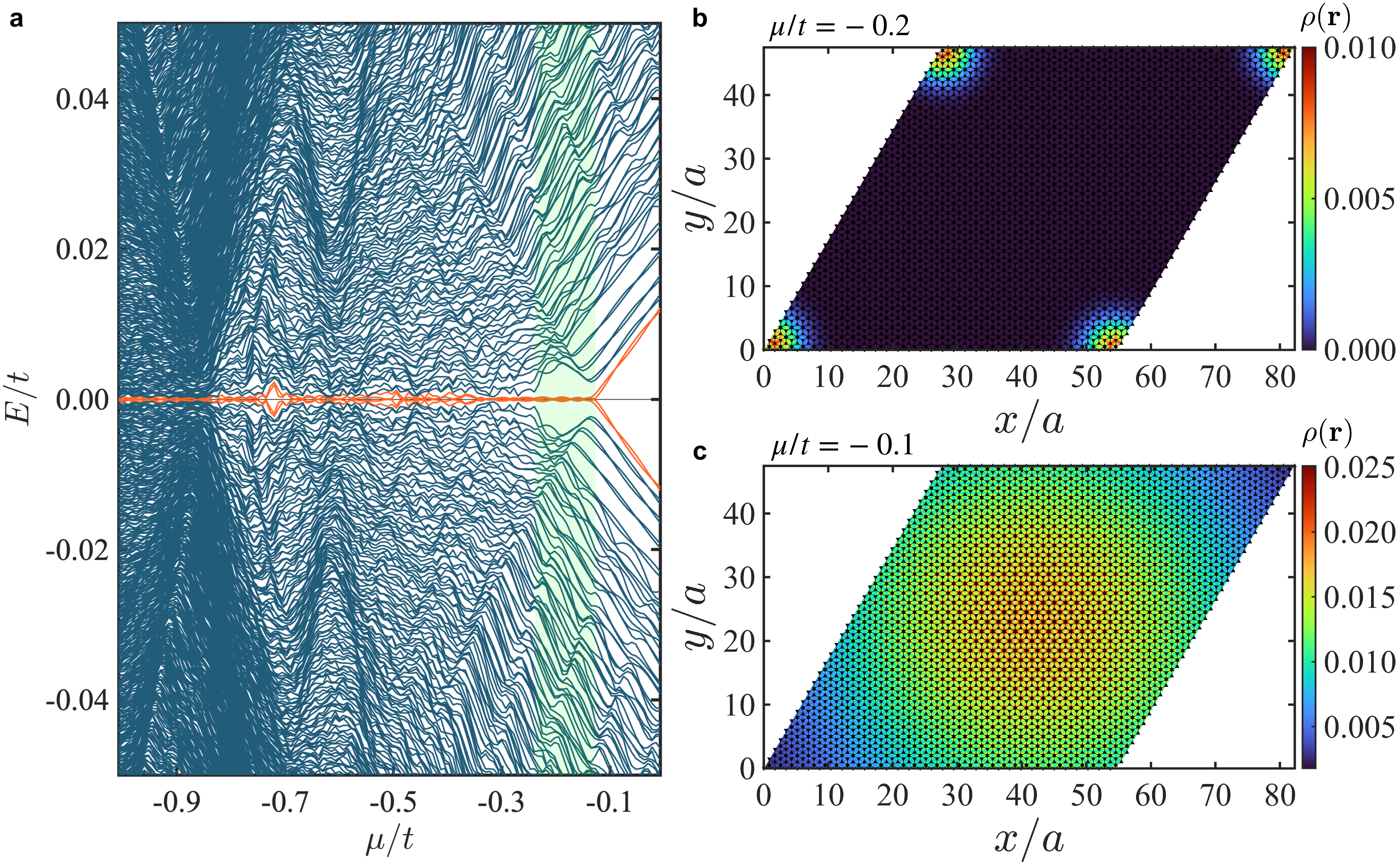,trim=0.0in 0.0in 0.0in 0.0in,clip=false, width=140mm}
\caption{{\bf Emergence of corner Majorana bound states (MBS).} {\bf {\small a}} Quasiparticle spectrum of a dice lattice of size $32\times32$ with open boundary conditions, with varying chemical potential $\mu$, revealing the range $-0.3t \lesssim \mu \lesssim -0.15t$, where two pairs of low-energy eigenstates come close to zero energy, while other eigenstates move to higher energies, creating a topological energy gap that protects the MBS. {\bf {\small b}}, {\bf {\small c}} Plots of the local density of states $\rho (\bf{r})$ (in arbitrary units) in the topological superconducting phase ($\mu \!=\!-0.2t$) and in the trivial superconducting phase ($\mu \!=\!-0.1t$). Other parameters: Rashba spin-orbit coupling strength $\lambda \!=\!0.1t$, external magnetic field amplitude $B_z\!=\!0.26t$, and hopping energy $t\!=\!1$. In {\bf {\small b}}, the localization of the MBS at the lattice corners indicates the second-order nature of the topological superconducting phase.}
\label{mu_var}
\vspace{-4mm}
\end{center}
\end{figure*}
\noindent where $t$ is the electron hopping amplitude, $i$ and $j$ are indices of different unit cells, $\alpha$ and $\beta$ represent indices of the three inequivalent sites within a unit cell, $\sigma=\uparrow,\downarrow$ labels the electron spin projection along the $z$ axis, $\langle \rangle$ represents nearest-neighbor (NN) sites, $\mu$ is the chemical potential, $\lambda$ is the strength of the Rashba spin-orbit coupling, $\mathbf{\hat{D}}_{ij}$ is the unit vector between unit cells $i$ and $j$, $\boldsymbol{\sigma}$ represents the Pauli matrices, $B_z$ is the strength of the magnetization field, the last two terms represent the onsite and non-local density-density attractive interactions with $U_0$ and $U_1$ as the strengths of the interactions, respectively, and $n_{i\alpha \sigma}=c_{i\alpha\sigma}^{\dagger}c_{i\alpha\sigma}$ is the electron density at the unit cell $i$, site $\alpha$ and spin $\sigma$. The interaction terms are treated at the mean-field level (see Methods section for details) and we obtain pairing amplitudes in different pairing channels as order parameters. For the self-consistent determination of the pairing amplitudes, we solve the Bogoliubov-de Gennes (BdG) equations, derived by performing the unitary transformation $c_{i\alpha\sigma}=\sum_{n}u_{i\alpha\sigma}^n\gamma_{n}^{\phantom{\dagger}}+v_{i\alpha\sigma}^{n*}\gamma_{n}^{\dagger}$ on the Hamiltonian~(\ref{Ham}), where $\gamma_{n}^{\phantom{\dagger}}$ is a fermionic annihilation operator in the $n^{\rm th}$ eigenstate, $u_{i\alpha\sigma}^n$ and $v_{i\alpha\sigma}^n$ are respectively the quasi-particle and quasi-hole amplitudes. We use, throughout this paper, lattice spacing $a\!=\!1$, hopping energy $t\!=\!1$, and attractive potentials $U_0\!=\!2t$ and $U_1\!=\!U_0/3$. We verified that a different choice for $U_0$ and $U_1$ does not change the conclusions presented here because the pairing amplitudes are calculated self-consistently. The triplet pairing amplitude is generated dynamically in the presence of Rashba spin-orbit coupling (broken inversion symmetry) and magnetic field (broken time-reversal symmetry)~\cite{PhysRevLett.87.037004,PhysRevB.97.214507}. Alternate routes to obtain spin-triplet topological pairing in similar systems include forward electron-phonon scattering which also suggests a robust equal-spin pairing~\cite{Li_CommPhys2023}.

\noindent  \textbf{\small \\Corner-localized MBS}\\
\indent We investigate the emergence of the zero-energy MBS by inspecting the quasiparticle spectrum, obtained by numerically solving the Hamiltonian~(\ref{Ham}) on a real lattice with open boundary conditions, while varying the chemical potential $\mu$. This procedure is repeated for many values of $\lambda$ and $B_z$, to search for signatures of the MBS in the quasiparticle spectrum. As shown in Fig.~\ref{mu_var}{\bf {\small a}}, at $(\lambda, B_z)\!=\!(0.1t, 0.26t)$ and  within the range $-0.3t \lesssim \mu \lesssim -0.15t$, two pairs of lowest-energy quasiparticle states remain close to zero energy while other low-energy levels move away towards higher energies, thus creating an energy gap. This energy gap provides topological protection to the zero-energy MBS, preventing them from hybridizing with the higher-energy ordinary quasiparticle states, in the presence of a local potential fluctuation. This energy gap, therefore, can also distinguish the corner MBS from other zero-energy non-Majorana states. To study the real-space localization of these zero-energy MBS in the two-dimensional dice lattice, in Fig.~\ref{mu_var}{\bf {\small b}}-{\bf {\small c}} we plot the local density of states, obtained via $\rho(\mathbf{r})\!=\! \sum_{\alpha,\sigma}(|u_{i\alpha\sigma}^n|^2+v_{i\alpha\sigma}^n|^2)$ with the index $n$ is taken to be the lowest-positive energy eigenstate. We use two values for the chemical potential: $\mu \!=\!-0.2t$, where the zero-energy states appear with a topological energy gap, and $\mu \!=\!-0.1t$, where the lowest-energy states are away from zero energy. At $\mu \!=\!-0.2t$, the lowest-energy eigenstate is localized at the four lattice corners, while at $\mu \!=\!-0.1t$ it is distributed inside the bulk. The corner-localized zero-energy states provide a strong indication of the appearance of the MBS, and hence of the induced second-order topological superconducting phase. An alternate route to obtain the corner MBS is to realize a second-order spin liquid phase~\cite{Vatsal_PRB2018,Principi_PRB2021}; we, however, restrict our discussions here to the case of second-order topological superconductivity. For lattices with a sublattice degree of freedom, such as dice, Lieb and kagom{\'e} lattices, the corner MBS can sensitively depend on the boundary termination as it can break some spatial symmetry~\cite{PhysRevB.97.205135,Kheirkhah_PRB2022}. It is interesting to note that the MBS at the diagonally-opposite lattice corners in our dice lattice are symmetric; this is because the opposite corners are related via mirror symmetry. It is, in fact, this mirror symmetry that protects the corner MBS in the dice lattice. In experimental realizations of these corner MBS, samples must be sufficiently clean so that quenched disorder does not damage the pairing and the subtle topological properties discussed here.\\

\noindent  \textbf{\small Pairing symmetry}\\
\indent The dice lattice has sites with coordination number both three and six, and this feature distinguishes it from the triangular and hexagonal lattices. The presence of these two types of sites determines the pairing symmetry in the superconducting state. The Rashba spin-orbit coupling also enforces its symmetry in the superconducting pairing. From the character table, shown in Table~\ref{table1}, one can notice that in this two-dimensional ${\cal D}_{3d}$ crystalline environment with broken both inversion symmetry (due to Rashba spin-orbit coupling) and time-reversal symmetry (due to the induced magnetization), the possible pairing symmetries arise from the E$_g$ \{$d_{xy}$, $d_{x^2-y^2}$\} (singlet pairing), and E$_u$ \{$p_x$, $p_y$\} (triplet pairing) irreducible representations.\\ 
\begin{figure}[t]
\begin{center}
\vspace{-0mm}
\epsfig{file=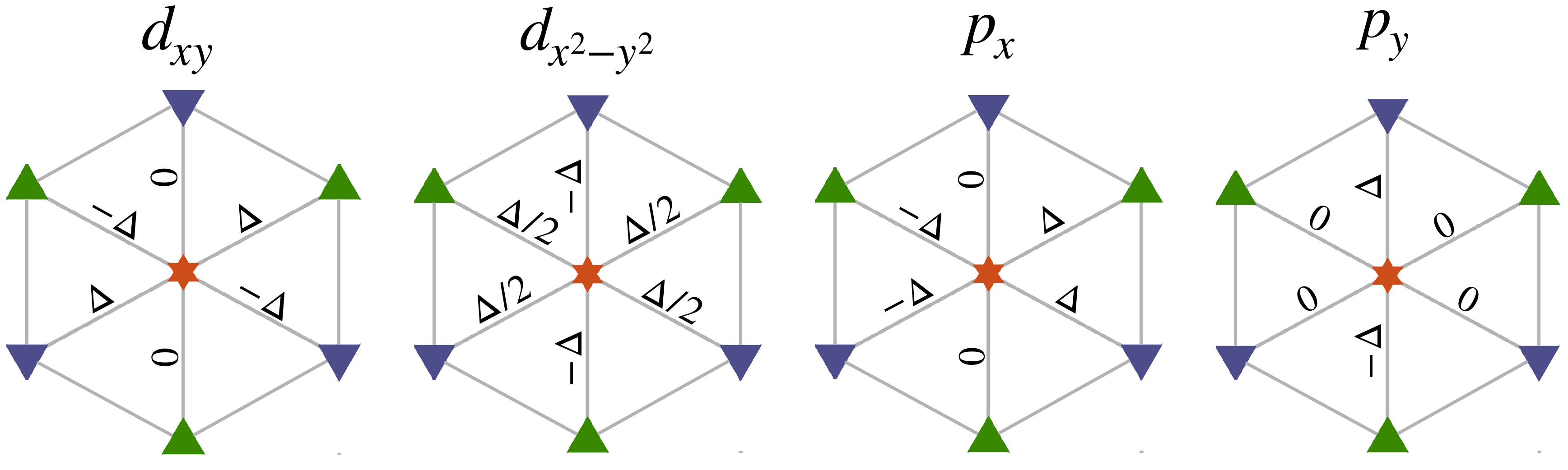,trim=0.0in 0.0in 0.0in 0.0in,clip=false, width=85mm}
\caption{{\bf Pairing symmetry in the dice lattice.} Possible pairing symmetries around a six-coordination site in the dice lattice with Rashba spin-orbit coupling and induced magnetization. $\Delta$ is the pairing amplitude used for the illustration of the amplitudes along different neighbors.}
\label{symmetry}
\vspace{-3mm}
\end{center}
\end{figure}


\vspace{-4mm}
\begin{center}
\begin{table}[h] \scriptsize
\begin{tabular}{|c|c|c|c|c|c|c|p{2cm}|p{2.2cm}|}
  \hline
  ${\cal D}$$_{3d}$ & E & 2C$_3$ & 3C$_2^{\prime}$ & ${\cal I}$  & 2S$_6$ & 3$\sigma_d$ & linear functions, rotations & quadratic functions \\
  \hline
  A$_{1g}$ & +1 & +1 & +1 & +1  & +1 & +1 & - & $x^2+y^2$, $z^2$ \\
    \hline 
    A$_{2g}$ & +1 & +1 & -1 & +1  & +1 & -1 & $R_z$ & - \\
    \hline
    E$_{g}$ & +2 & -1 & 0 & +2  & -1 & 0 & $R_x$, $R_y$ & $x^2$-$y^2$, $xy$, $xz$, $yz$ \\
    \hline
    A$_{1u}$ & +1 & +1 & +1 & -1  & -1 & -1 & - & - \\
    \hline
    A$_{2u}$ & +1 & +1 & -1 & -1  & -1 & +1 & $z$ & - \\
    \hline
    E$_{u}$ & +2 & -1 & 0 & -2  & +1 & 0 & $x$, $y$ & - \\
    \hline
\end{tabular}
  \caption{Character table for the ${\cal D}$$_{3d}$ point group. $g$ and $u$ represent, respectively, the symmetric and anti-symmetric wave functions with respect to the inversion center.}
\label{table1}
\end{table}
\vspace{-8mm}
\end{center}

\begin{figure*}[t]
\begin{center}
\vspace{0mm}
\epsfig{file=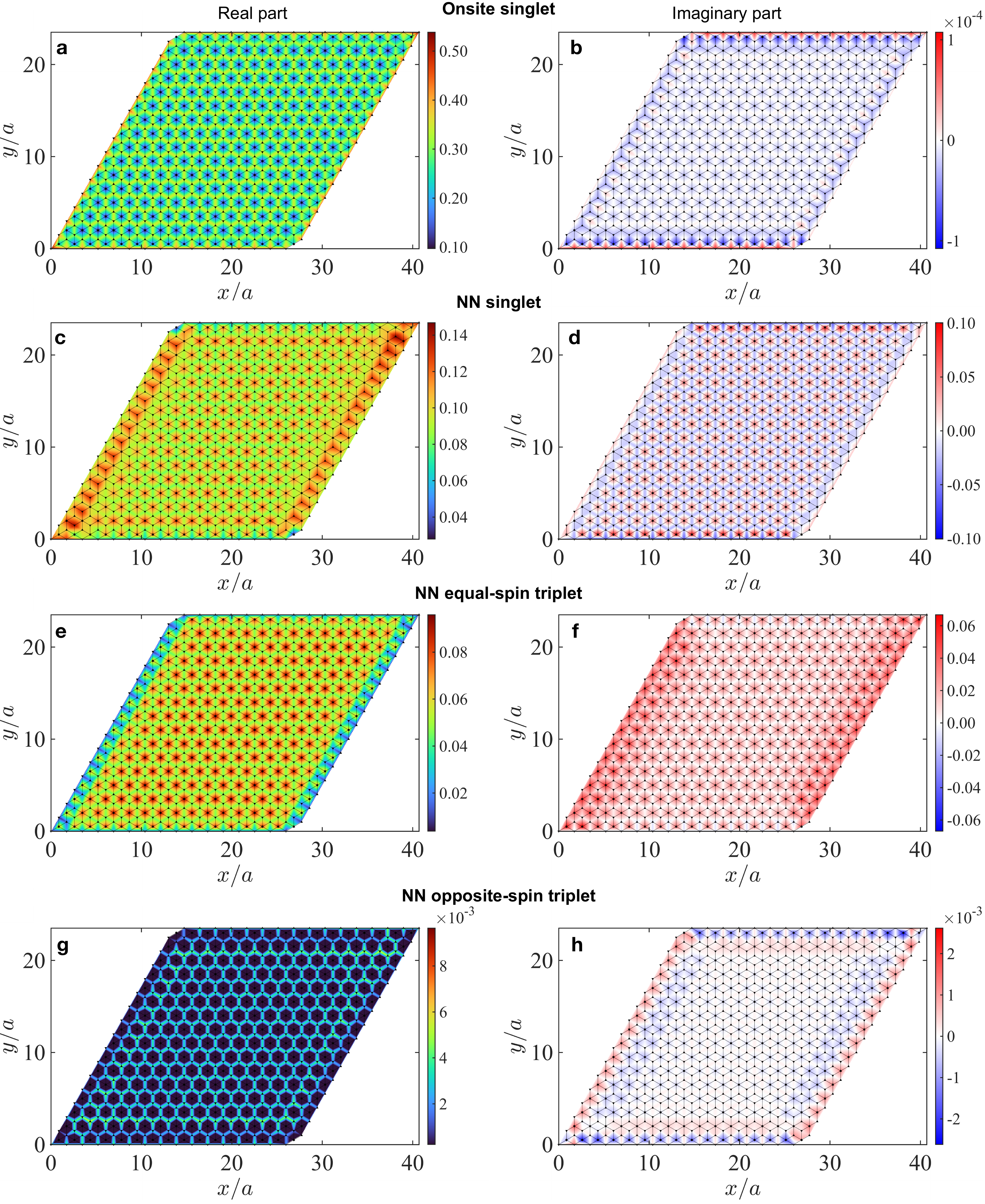,trim=0.0in 0.0in 0.0in 0.0in,clip=false, width=150mm}
\caption{{\bf Real-space profile of pairing amplitudes.} Real and imaginary parts of the pairing amplitudes for all possible pairing channels: {\bf {\small a}}-{\bf {\small b}} onsite singlet,  {\bf {\small c}}-{\bf {\small d}} nearest-neighbor (NN) singlet, {\bf {\small e}}-{\bf {\small f}} NN equal-spin ($\uparrow \uparrow$) triplet, and {\bf {\small g}}-{\bf {\small h}} NN opposite-spin triplet, on a dice lattice of size 16$\times$16 with open boundary conditions. Parameters used: Rashba spin-orbit coupling strength $\lambda \!=\!0.1t$, external magnetic field amplitude $B_z\!=\!0.26t$, chemical potential $\mu \!=\!-0.2t$, and hopping energy $t\!=\!1$.}
\label{pairings}
\vspace{-4mm}
\end{center}
\end{figure*}

These possible pairing channels are shown schematically in Fig.~\ref{symmetry}. Mixing of the singlet and triplet components is allowed by the broken structural inversion symmetry in the discussed oxide trilayers~\cite{PhysRevLett.87.037004}. Therefore, a linear combination of these four types of pairing symmetry is stabilized. Figure~\ref{pairings} shows the profiles of the pairing amplitudes on the dice lattice at the same set of parameters where the corner MBS are found. The imaginary components of the nearest-neighbor (NN) pairing amplitudes are nonzero, implying a chiral mixed-parity topological superconducting state. The real part of the onsite singlet pairing amplitude Re($\Delta_{i}^{\rm s, On}$) (Fig.~\ref{pairings}{\bf {\small a}}) clearly reveals a difference between the three and six coordination sites. The imaginary part of the onsite singlet pairing amplitude Im($\Delta_{i}^{\rm s, On}$) (Fig.~\ref{pairings}{\bf {\small b}}) vanishes inside the bulk as expected, but it has a small finite value at the edges only in the presence of a finite Rashba spin-orbit coupling. While the onsite singlet pairing amplitude Re($\Delta_{i}^{\rm s, On}$) at the six-coordination sites is slightly smaller than that at the three-coordination sites, the NN singlet pairing amplitude Re($\Delta_{i}^{\rm s, NN}$)  (Fig.~\ref{pairings}{\bf {\small c}}) shows the opposite behavior. On the other hand, the imaginary part of the NN singlet pairing amplitude Im($\Delta_{i}^{\rm s, NN}$) (Fig.~\ref{pairings}{\bf {\small d}}), at the three and six -coordination sites are of different magnitudes and signs. The real part of the NN equal-spin triplet pairing amplitude Re($\Delta_{i,\sigma \sigma}^{\rm t, NN}$) (Fig.~\ref{pairings}{\bf {\small e}}) also has a larger value at the six-coordination sites than the three-coordination ones, while its imaginary part Im($\Delta_{i,\sigma \sigma}^{\rm t, NN}$) (Fig.~\ref{pairings}{\bf {\small f}}) vanishes at the three-coordination sites. The real part of the NN opposite-spin triplet pairing amplitude Re($\Delta_{i,\uparrow \downarrow}^{\rm t, NN}$) (Fig.~\ref{pairings}{\bf {\small g}}) is an order of magnitude smaller than the equal-spin triplet pairing amplitude and it vanishes completely at the six-coordination sites. The imaginary part Im($\Delta_{i,\uparrow \downarrow}^{\rm t, NN}$) (Fig.~\ref{pairings}{\bf {\small h}}) vanishes at all sites except those near the boundaries. The slight variation in the pairing amplitudes near the corners and edges of the lattice is due to the considered open boundary conditions. The above results confirms that odd-parity, equal-spin pairing in the triplet channel is favored over the opposite-spin one due to parity fluctuations in the presence of Rashba spin-orbit coupling and a time-reversal symmetry-breaking Zeeman exchange field~\cite{PhysRevLett.115.207002}. \\

\begin{figure*}[t]
\begin{center}
\vspace{0mm}
\epsfig{file=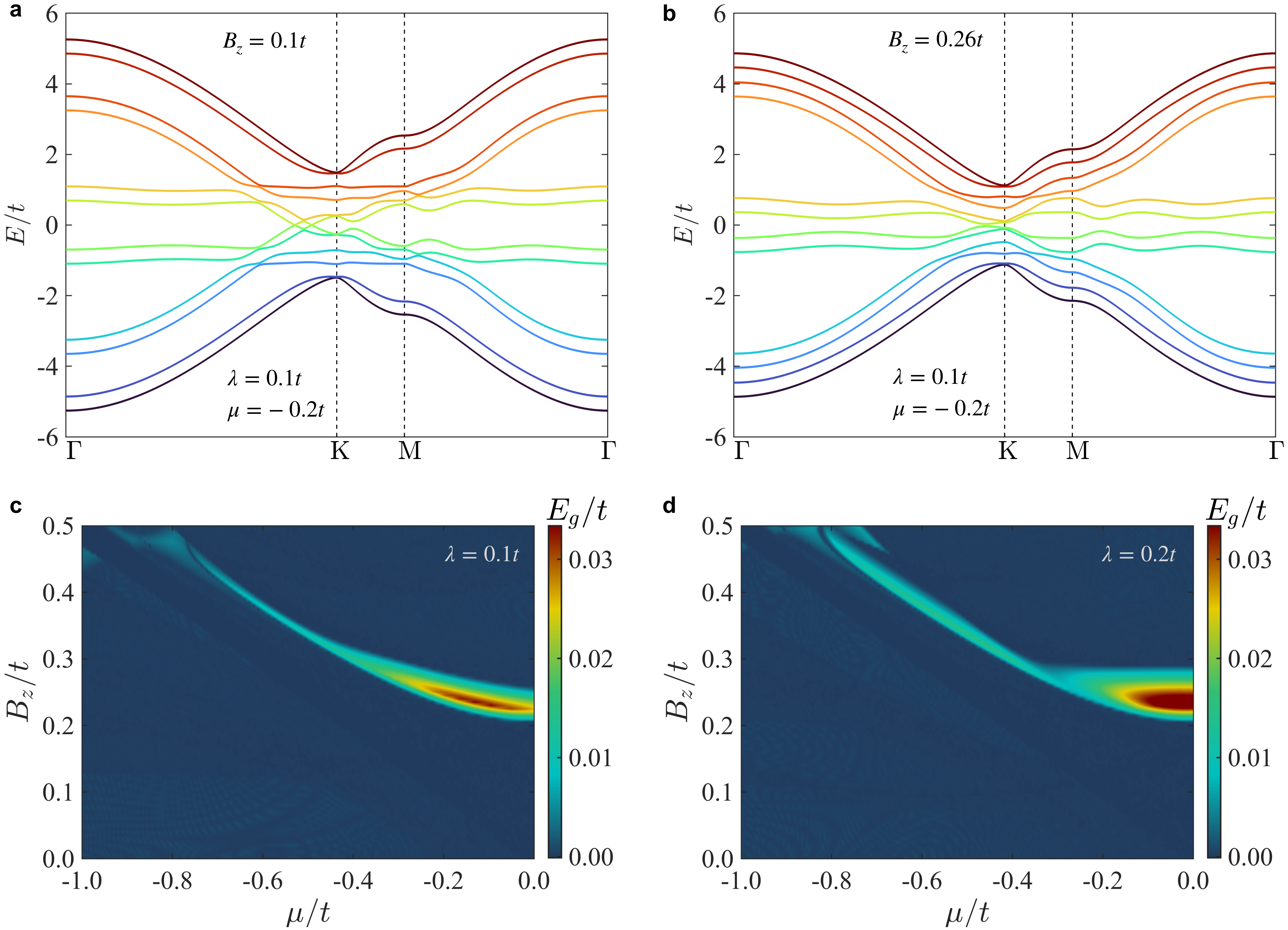,trim=0.0in 0.0in 0.0in 0.0in,clip=false, width=170mm}
\caption{{\bf Quasiparticle bands and excitation energy gap.} {\bf {\small a}}-{\bf {\small b}} Quasiparticle bands along the momentum path $\Gamma (0,0)-{\rm K}(\frac{4\pi}{3\sqrt{3}},0)-{\rm M}(\frac{\pi}{\sqrt{3}},\frac{\pi}{3})-\Gamma$, in the trivial phase (zero excitation gap at a magnetic field strength $B_z \!=\! 0.1t$) and topological superconducting phase (a finite excitation gap at $B_z \!=\! 0.26t$). {\bf {\small c}}-{\bf {\small d}} Excitation gap $E_g$, shown by the colorbar, plotted in the plane of $B_z$ and chemical potential $\mu$ at two values of the Rashba spin-orbit coupling strengths $\lambda=0.1t$, $\lambda=0.2t$. The hopping energy is $t\!=\!1$.}
\label{kspace_bands}
\vspace{-4mm}
\end{center}
\end{figure*}

\noindent  \textbf{\small Topological superconducting transition}\\
\indent The transition to the second-order topological superconducting phase can be understood by inspecting the quasiparticle band dispersion in momentum space, obtained by diagonalizing the following BdG Hamiltonian at 
wavevector $\mathbf{k}\equiv(k_x,k_y)$
\begin{align}
{\cal H}_{\rm BdG}(\mathbf{k})\!=\!
\begin{pmatrix}
\Psi_{\mathbf{k}}^{\dagger} & \Psi_{-\mathbf{k}}
\end{pmatrix}
\begin{pmatrix}
{\cal H}_e   & {\cal H}_{\Delta} \\
{\cal H}_{\Delta}^* & {\cal H}_h
\end{pmatrix}
\begin{pmatrix}
\Psi_{\mathbf{k}} \\
\Psi_{-\mathbf{k}}^{\dagger}
\end{pmatrix},
\label{H_k}
\end{align}
where $\Psi_{\mathbf{k}}=[c_{\mathbf{k}1\uparrow}, c_{\mathbf{k}2\uparrow}, c_{\mathbf{k}3\uparrow},c_{\mathbf{k}1\downarrow}, c_{\mathbf{k}2\downarrow}, c_{\mathbf{k}3\downarrow}]^T$; 1, 2, 3 denote the three inequivalent sites within a unit cell; ${\cal H}_e$, ${\cal H}_h$ and ${\cal H}_{\Delta}$ are the matrices representing, respectively, the electron, the hole, and the pairing sectors of the Hamiltonian, described in the Methods section. We show the quasiparticle spectrum, in Fig.~\ref{kspace_bands}{\bf a}-{\bf b}, at two values of the magnetic field, in the vicinity of the parameter regime in which the corner MBS were found in the above real-space analysis. The two lowest-energy pairs of the quasiparticle bands close the gap near the $\rm K$ point along the $\Gamma$-$\rm K$ direction for most of the parameter regime, as shown in Fig.~\ref{kspace_bands}{\bf a} for $B_z\!=\!0.1t$. However, a small gap is opened, indicating possible topological superconducting transition, when the field is increased, as shown in Fig.~\ref{kspace_bands}{\bf b} for $B_z\!=\!0.26t$. We, therefore, use the quasiparticle excitation gap $E_g= $~min$(E_{1}(\mathbf{k}))$, defined as the minimum of the 1$^{\rm st}$ positive (or negative) quasiparticle band, as a diagnostic tool to locate the topological superconducting state. In Fig.~\ref{kspace_bands}{\bf c} and {\bf d}, we show this excitation gap $E_g$ in the plane of $\mu$ and $B_z$, for two values of the Rashba spin-orbit coupling strength $\lambda$. The plots show the appearance of a well-defined parameter regime, bounded by two critical values of $B_z$ or $\mu$, with a finite $E_g$. The corner MBS were found in the above analysis in this parameter regime with a small quasiparticle excitation gap. The identification of a topological invariant for the discussed second-order topological superconductivity in the dice lattice requires careful consideration of the available symmetries and the fractional charges at the lattice corners, as derived for higher-order topological insulating systems~\cite{Hughes_Science2017}; we leave such a possibility for future studies.

\vspace{-1em}
\section*{Conclusion}
\vspace{-1em}
To summarize, we showed that topological flat bands with Chern number 2 in the dice lattice with attractive interaction among electrons harbor a second-order topological superconducting phase. A signature of this exotic topological phase is revealed by the presence of the MBS at the lattice corners. Analogies between the topological superconductivity in flat bands, as found here, and the quantum-Hall insulator/superconductor interfaces can be drawn. Theoretically, it is known that a quantum Hall state with Chern number 1, in proximity to a fully gapped $s$-wave superconductor, generates a topological first-order superconducting phase~\cite{PhysRevB.82.184516,PhysRevResearch.2.023063}. Likewise, the fractionalized MBS, \textit{i.e.} some realizations of the parafermions, have been proposed in fractional quantum Hall states when in proximity to an $s$-wave superconductor~\cite{Clarke2013,PhysRevX.4.031009,PhysRevResearch.2.013232}. These findings establish a close connection between the topology of the normal state and the nature of the induced topological superconductivity. Topological flat bands with higher Chern numbers are found not only in the dice lattice, but also in kagom{\'e} and Lieb lattices~\cite{PhysRevB.104.235115,Okamoto_CommPhys2022}. Other than the examples of a few-layer graphene and a transition-metal-oxide trilayer, another candidate compound is CsV$_3$Sb$_5$~\cite{PhysRevLett.125.247002, Zhao2021, HU2022495}, where lattice geometry, flat-band topology and superconductivity can produce Majorana states such as those discussed here. Hence, we expect that future research will unveil topological superconductivity in a variety of compounds that exhibit topological flat bands. Furthermore, the superconducting transition temperature is proportional to the density of states at the Fermi level which is large for these flat-band systems. Therefore, when looking forward, topological flat-bands with higher Chern numbers provide an opportunity to search for higher-order topological superconductivity at high temperatures.

\section*{Methods}
\noindent  \textbf{\small Calculation of pairing amplitudes\\}
\indent The  attractive interaction terms in the Hamiltonian~(\ref{Ham}) are decomposed into different pairing channels (singlet and triplet, onsite and nearest-neighbor) and the resulting mean-field Hamiltonian for these two interaction terms is given by
{\small
\begin{align}
\vspace{-4mm}
{\cal H}_{MF}\!&=\! \sum_{i,\alpha}(\Delta_{ii}^{\alpha\alpha} c_{i\alpha\uparrow}^{\dagger} c_{i\alpha\downarrow}^{\dagger}+\text{H.c.}) \nonumber \\
&+\!\frac{1}{2}\!\sum_{\langle ij \rangle,\alpha,\beta,\sigma,\sigma^{\prime}}\!(\Delta_{ij\sigma\sigma^{\prime}}^{\alpha\beta} c_{i\alpha\sigma}^{\dagger} c_{j\beta\sigma^{\prime}}^{\dagger}\!+\!\text{H.c.}) \nonumber \\
&+\sum_{i,\alpha,\sigma}\Gamma_{ii\sigma}^{\rm H} c_{i\alpha\sigma}^{\dagger} c_{i\alpha\sigma}
+\frac{1}{2}\sum_{\langle ij \rangle,\alpha,\beta,\sigma}\Gamma_{ij}^{\rm H} c_{i\alpha\sigma}^{\dagger} c_{i\alpha\sigma} \nonumber \\
&-\frac{1}{2}\sum_{\langle ij \rangle,\alpha,\beta,\sigma,\sigma^{\prime}}\Gamma_{ij\sigma\sigma^{\prime}}^{\rm F} c_{i\alpha\sigma}^{\dagger} c_{j\beta\sigma^{\prime}},
\label{H_MF}
\vspace{-4mm}
\end{align} 
}
\noindent where the on-site and off-site pairing amplitudes $\Delta_{ii}^{\alpha\alpha}$, $\Delta_{ij\sigma\sigma^{\prime}}^{\alpha\beta}$, the on-site Hartree potential $\Gamma_{ii\sigma}^{\rm H}$, the off-site Hartree potential $\Gamma_{ij}^{\rm H}$ and the   Fock potential $\Gamma_{ij\sigma\sigma^{\prime}}^{\rm F}$ are obtained self-consistently via the following relations
{\small
\begin{align}
\vspace{-4mm}
&\Delta_{ii}^{\alpha\alpha}=-U_0 \langle c_{i\alpha\downarrow}c_{i\alpha\uparrow} \rangle \nonumber \\
&\Delta_{ij\sigma\sigma^{\prime}}^{\alpha\beta}=-\frac{U_1}{2} \langle c_{i\alpha\sigma} c_{j\beta\sigma^{\prime}} \rangle \nonumber \\
&\Gamma_{ii\sigma}^{\rm H}=-U_0 \langle c_{i\alpha\sigma}^{\dagger} c_{i\alpha\sigma} \rangle \nonumber \\
&\Gamma_{ij}^{\rm H}=-U_1 \sum_{\sigma} \langle c_{j\alpha\sigma}^{\dagger} c_{j\alpha\sigma} \rangle \nonumber \\
&\Gamma_{ij\sigma\sigma^{\prime}}^{\rm F}=-\frac{U_1}{2} \langle c_{i\alpha\sigma}^{\dagger} c_{j\beta\sigma^{\prime}} \rangle
\label{HatreeFock}
\vspace{-4mm}
\end{align} 
}
The total Hamiltonian is then diagonalized using the BdG transformation $c_{i\alpha\sigma}\!=\!\sum_{n}u_{i\alpha\sigma}^n\gamma_{n}+v_{i\alpha\sigma}^{n*}\gamma_{n}^{\dagger}$, where $\gamma_{n}$ is a fermionic annihilation operator at the $n^{\rm th}$ eigenstate, $u_{i\alpha\sigma}^n$ and $v_{i\alpha\sigma}^n$ are  the quasi-particle and quasi-hole amplitudes, respectively. The quasi-particle and quasi-hole amplitudes are obtained by solving the BdG equations $\sum_{j}{\cal H}_{ij}\psi_{j}^n \!=\!\epsilon_n\psi_{i}^n$, where $\psi_{i}^n\!=\![u_{i\alpha\uparrow}^n, u_{i\alpha\downarrow}^n, v_{i\alpha\uparrow}^n, v_{i\alpha\downarrow}^n]^T$ with $u_{i\alpha\uparrow}^n=[u_{i1\uparrow}^n,u_{i2\uparrow}^n,u_{i3\uparrow}^n]$ and similarly for other components, while $\epsilon_n$ is the energy eigenvalue of the $n^{\rm th}$ eigenstate. The self-consistency iterations continue until all the pairing amplitudes converge at all lattice sites, within a tolerance of 10$^{-8}$. Finally, the following order parameters were calculated from the converged eigenvalues and eigenvectors:

\vspace{-4mm}
{\small
\begin{align}
&\text{ On-site singlet: }\Delta_{i}^{\rm s, On}=-U_0 \langle c_{i\alpha\downarrow}c_{i\alpha\uparrow} \rangle \nonumber \\
&\text{ NN singlet: }\Delta_{i}^{\rm s, NN}=-\frac{U_1}{2N_n}\sum_{\langle j \alpha \beta \rangle} \langle c_{i\alpha\downarrow}c_{j\beta\uparrow} -  c_{i\alpha\uparrow}c_{j\beta\downarrow}\rangle \nonumber \\
&\text{ NN equal-spin triplet: }\Delta_{i,\sigma \sigma}^{\rm t, NN}=-\frac{U_1}{N_n}\sum_{\langle j \alpha \beta \rangle} \langle c_{i\alpha\sigma}c_{j\beta\sigma} \rangle \\
&\text{ NN opposite-spin triplet: }\nonumber \\
&\hspace{15 mm} \Delta_{i,\uparrow \downarrow}^{\rm t, NN}\!=\!-\frac{U_1}{2N_n}\!\sum_{\langle j \alpha \beta \rangle}\! \langle c_{i\alpha\downarrow}c_{j\beta\uparrow} \!+\!  c_{i\alpha\uparrow}c_{j\beta\downarrow}\rangle
\nonumber
\end{align}
}
where $N_n$ denotes the number of NN.\\

\begin{figure}[h]
\begin{center}
\vspace{-2mm}
\epsfig{file=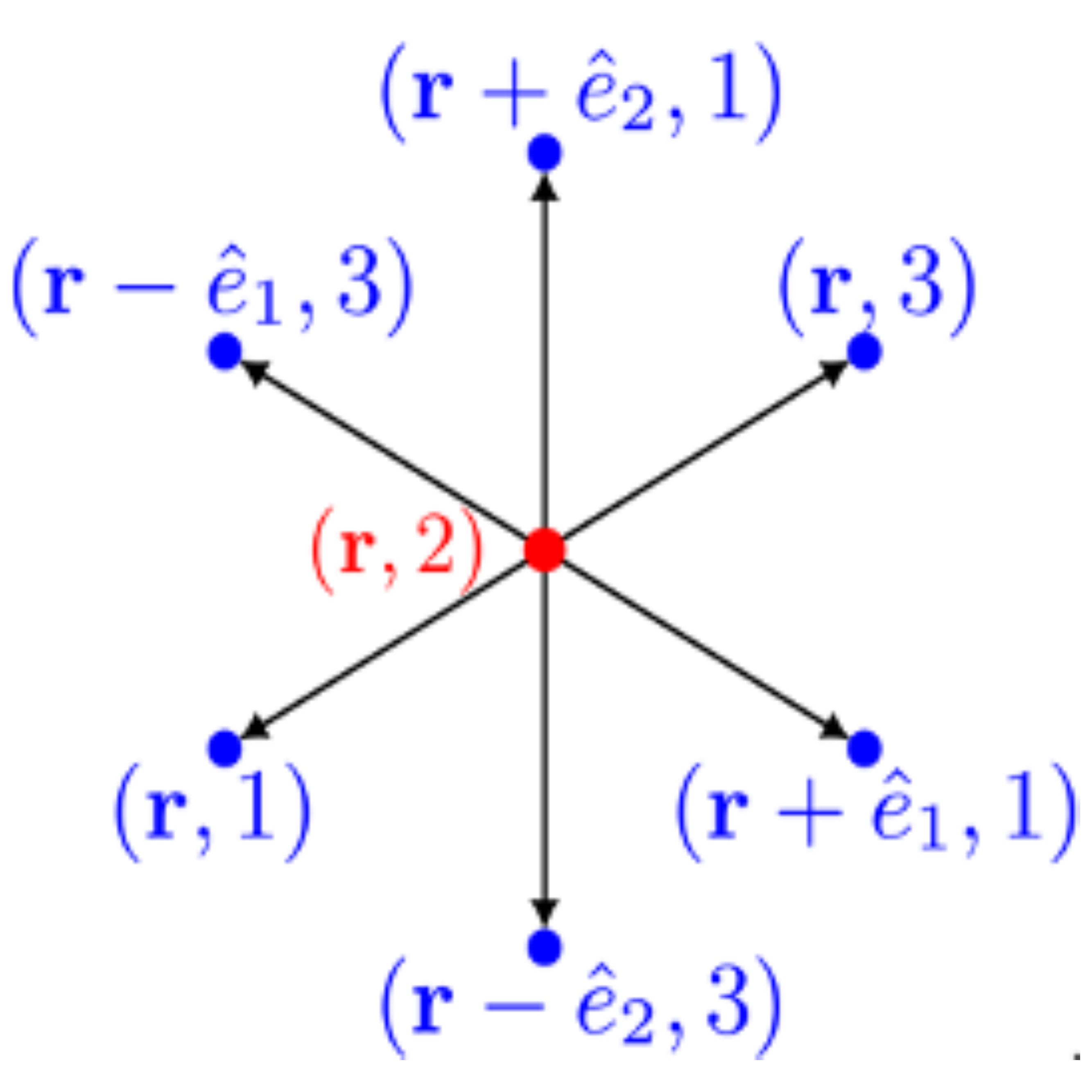,trim=0.0in 0.0in 0.0in 0.0in,clip=false, width=45mm}
\caption{Notation for the coordinates in the pairing terms in Eq.~(\ref{gaps}). Red and blue dots represent, respectively, the six and three -coordination sites of the dice lattice.}
\label{coords}
\vspace{-2mm}
\end{center}
\end{figure}

\noindent  \textbf{\small \\Momentum-space Hamiltonian\\}
\indent The Hamiltonian~(\ref{H_k}) is expressed in the basis $\Psi_{\mathbf{k}}=[c_{\mathbf{k}1\uparrow}, c_{\mathbf{k}2\uparrow}, c_{\mathbf{k}3\uparrow},c_{\mathbf{k}1\downarrow}, c_{\mathbf{k}2\downarrow}, c_{\mathbf{k}3\downarrow}]^T$, where 1, 2, 3 denote the three inequivalent sites within a unit cell, and is given by
{\scriptsize
\begin{align}
&{\cal H}_{e}(\mathbf{k})\!=\! \nonumber \\
&\begin{pmatrix}
-B_z-\mu  & -t\gamma_{\mathbf{k}}^* & 0 & 0 &  -i\lambda \gamma_{\mathbf{k+}}^* & 0\\
-t\gamma_{\mathbf{k}}  & -B_z-\mu & -t\gamma_{\mathbf{k}}^* & i\lambda \gamma_{\mathbf{k-}} & 0 &  i\lambda \gamma_{\mathbf{k+}}^*\\
0  & -t\gamma_{\mathbf{k}} & -B_z-\mu & 0 &  -i\lambda\gamma_{\mathbf{k-}} & 0\\
0  & -i\lambda\gamma_{\mathbf{k-}}^* & 0 & B_z-\mu &  -t\gamma_{\mathbf{k}}^* & 0\\
i\lambda \gamma_{\mathbf{k+}}  & 0 & i\lambda\gamma_{\mathbf{k-}}^* & -t\gamma_{\mathbf{k}} & B_z-\mu &-t\gamma_{\mathbf{k}}^*\\
0  & -i\lambda\gamma_{\mathbf{k+}} & 0 & 0 &  -t\gamma_{\mathbf{k}} & B_z-\mu
\end{pmatrix},
\end{align}
}
\noindent where $\gamma_{\mathbf{k}}\!=\!1\!+\!e^{ik_1}\!+\!e^{ik_2}$, $\gamma_{\mathbf{k\pm}}\!=\!1+e^{ik_1\pm2\pi/3}+e^{ik_2\pm4\pi/3}$, $k_{1,2}\!=\!\mathbf{k}\cdot \mathbf{e}_{1,2}$, and $\mathbf{e}_{1,2}$ are the lattice translational vectors, given by $\mathbf{e}_{1}\! \equiv \!(\sqrt 3,~0)$ and  $\mathbf{e}_{2}\! \equiv \!(\sqrt 3/2,~3/2)$.
The topological flat bands in Fig.~\ref{dice_flatbands}{\small{\bf B}} are obtained by diagonalizing ${\cal H}_{e}(\mathbf{k})$ at $t\!=\!1$, $\mu \!=\!0$, $\lambda \!=\!0.3t$, and $B_z\!=\!0.4t$. The hole part of the Hamiltonian~(\ref{H_k}) is given by ${\cal H}_{h}(\mathbf{k})\!=\![-{\cal H}_{e}(-\mathbf{k})]^T$, and the pairing part is given by
{\scriptsize
\begin{align}
&{\cal H}_{\Delta}(\mathbf{k})\!=\! \nonumber \\
&\begin{pmatrix}
0  &  -\Delta_{12}^{\uparrow\uparrow}(-\mathbf{k}) & 0 & \Delta_{11}^{\rm s} & -\Delta_{12}^{s/t}(-\mathbf{k}) & 0\\
-\Delta_{12}^{\uparrow\uparrow}(\mathbf{k})  & 0 & \Delta_{32}^{\uparrow\uparrow}(\mathbf{k}) & -\zeta \Delta_{12}^{s/t}(\mathbf{k}) & \Delta_{22}^{\rm s}  & -\zeta \Delta_{32}^{s/t}(\mathbf{k})\\
0  & -\Delta_{32}^{\uparrow\uparrow}(-\mathbf{k}) & 0 & 0 & -\Delta_{32}^{s/t}(-\mathbf{k}) & \Delta_{33}^{\rm s} \\
-\Delta_{11}^{\rm s} & \zeta \Delta_{12}^{s/t}(-\mathbf{k}) & 0 & 0 & -\Delta_{12}^{\downarrow\downarrow}(-\mathbf{k}) & 0\\
\Delta_{12}^{s/t}(\mathbf{k}) & -\Delta_{22}^{\rm s}  & \Delta_{32}^{s/t}(\mathbf{k}) & -\Delta_{12}^{\downarrow\downarrow}(\mathbf{k}) & 0 &  \Delta_{32}^{\downarrow\downarrow}(\mathbf{k})\\
0 & \zeta \Delta_{32}^{s/t}(-\mathbf{k}) & -\Delta_{33}^{\rm s} & 0 & -\Delta_{32}^{\downarrow\downarrow}(-\mathbf{k}) & 0 \\
\end{pmatrix},
\end{align}
}
\noindent where $\Delta_{\alpha\alpha}^{\rm s}$ ($\alpha \!=\!1,2,3$) represents the onsite singlet pairing amplitude at site index $\alpha$, $\zeta=+1 (-1)$ for singlet (triplet) pairing, $\Delta_{12}(\mathbf{k})$ and $\Delta_{32}(\mathbf{k})$ are the NN pairing amplitudes and expressed below.
{\small
\begin{align}
&\Delta_{12}(\mathbf{k})=\Delta_{\mathbf{r},1;\mathbf{r},2}+\Delta_{\mathbf{r+\hat{e}_1},1;\mathbf{r},2}e^{-ik_1}+\Delta_{\mathbf{r+\hat{e}_2},1;\mathbf{r},2}e^{-ik_2} \nonumber \\
&\Delta_{32}(\mathbf{k})=\Delta_{\mathbf{r},3;\mathbf{r},2}+\Delta_{\mathbf{r-\hat{e}_1},3;\mathbf{r},2}e^{ik_1}+\Delta_{\mathbf{r-\hat{e}_2},1;\mathbf{r},2}e^{ik_2}
\label{gaps}
\end{align}
}
\noindent Here the real-space pairing amplitudes are defined based on the NN hopping between the three inequivalent sites, as described in Fig.~\ref{coords}.
Based on the above notation, and pairing symmetries described in Fig.~\ref{symmetry}, one can collect the amplitudes for $d_{xy}$, $d_{x^2-y^2}$, $p_{x}$ and $p_{y}$ -wave pairing channels. These are given by\\

\noindent $d_{xy}$-\text{wave}:
\vspace{-2mm}
{\small
\begin{align}
&\Delta_{12}(\mathbf{k})=\Delta_{d_{xy}}(1-e^{-ik_1}) \nonumber \\
&\Delta_{32}(\mathbf{k})=\Delta_{d_{xy}}(1-e^{ik_1}) 
\end{align}
}
\noindent $d_{x^2-y^2}$-\text{wave}:
\vspace{-2mm}
{\small
\begin{align}
&\Delta_{12}(\mathbf{k})=\frac{\Delta_{d_{x^2-y^2}}}{2}(1+e^{-ik_1}-2e^{-ik_2}) \nonumber \\
&\Delta_{32}(\mathbf{k})=\frac{\Delta_{d_{x^2-y^2}}}{2}(1+e^{ik_1}-2e^{ik_2}) 
\end{align}
}
\noindent $p_{x}$-\text{wave}:
\vspace{-2mm}
{\small
\begin{align}
&\Delta_{12}(\mathbf{k})=\Delta_{p_{x}}(1+e^{-ik_1}) \nonumber \\
&\Delta_{32}(\mathbf{k})=\Delta_{p_{x}}(1-e^{ik_1}) 
\end{align}
}
\noindent $p_{y}$-\text{wave}:
\vspace{-2mm}
{\small
\begin{align}
&\Delta_{12}(\mathbf{k})=\Delta_{p_{y}}(1+e^{-ik_2}) \nonumber \\
&\Delta_{32}(\mathbf{k})=\Delta_{p_{y}}(1-e^{ik_2}) 
\end{align}
}
Taking cue from the results of the real-space analysis, presented in Fig.~\ref{pairings}, at parameters $\lambda \!=\!0.1t$, $B_z\!=\!0.26t$, and $\mu \!=\!-0.2t$, we use the following set of gap parameters $\Delta_{11}^{\rm s}=\Delta_{33}^{\rm s}=0.4t$, $\Delta_{22}^{\rm s}=0.2t$, $\Delta_{d_{xy}}=\Delta_{d_{x^2-y^2}}=0.1t$, $\Delta_{p_{x}}=\Delta_{p_{y}}=0.1t$ (for $\uparrow\uparrow$), $\Delta_{p_{x}}=\Delta_{p_{y}}=0.02t$ (for $\downarrow\downarrow$). All these pairing amplitudes are used additively to construct the pairing Hamiltonian ${\cal H}_{\Delta}(\mathbf{k})$.


 
\section*{Acknowledgements}
\vspace{-1em}
This work was supported by the U.S. Department of Energy, Office of Science, Basic Energy Sciences, Materials Sciences and Engineering Division. NM acknowledges the National Supercomputing Mission for providing computing resources of Param Ganga at the Indian Institute of Technology Roorkee, which is implemented by C-DAC and supported by the Ministry of Electronics and Information Technology and Department of Science and Technology, Government of India.

\section*{Author contributions}
\vspace{-1em}
NM planned the work, performed numerical calculations and wrote the manuscript with inputs from all coauthors. RS provided inputs in setting up the momentum-space Hamitonian. SO  provided inputs in the analysis of the topological superconducting phase. ED provided inputs in the analysis of the interaction terms.




\begin{thebibliography}{10}
\vspace{-1em}
\expandafter\ifx\csname url\endcsname\relax
  \def\url#1{\texttt{#1}}\fi
\expandafter\ifx\csname urlprefix\endcsname\relax\def\urlprefix{URL: }\fi
\providecommand{\bibinfo}[2]{#2}
\providecommand{\eprint}[2][]{\url{#2}}

\bibitem{Schindler_SciAdv2018}
\bibinfo{author}{Schindler, F.} \textit{et~al.}
\newblock \bibinfo{title}{Higher-order topological insulators}.
\newblock \textit{\bibinfo{journal}{Sci. Adv.}} \textbf{\bibinfo{volume}{4}},
  \bibinfo{pages}{eaat0346} (\bibinfo{year}{2018}).
\newblock
  \urlprefix\url{https://www.science.org/doi/abs/10.1126/sciadv.aat0346}.

\bibitem{PhysRevLett.119.246401}
\bibinfo{author}{Langbehn, J.}, \bibinfo{author}{Peng, Y.},
  \bibinfo{author}{Trifunovic, L.}, \bibinfo{author}{von Oppen, F.} \&
  \bibinfo{author}{Brouwer, P.~W.}
\newblock \bibinfo{title}{Reflection-symmetric second-order topological
  insulators and superconductors}.
\newblock \textit{\bibinfo{journal}{Phys. Rev. Lett.}}
  \textbf{\bibinfo{volume}{119}}, \bibinfo{pages}{246401}
  (\bibinfo{year}{2017}).
\newblock
  \urlprefix\url{https://link.aps.org/doi/10.1103/PhysRevLett.119.246401}.

\bibitem{PhysRevLett.120.026801}
\bibinfo{author}{Ezawa, M.}
\newblock \bibinfo{title}{Higher-order topological insulators and semimetals on
  the breathing kagom{\'e} and pyrochlore lattices}.
\newblock \textit{\bibinfo{journal}{Phys. Rev. Lett.}}
  \textbf{\bibinfo{volume}{120}}, \bibinfo{pages}{026801}
  (\bibinfo{year}{2018}).
\newblock
  \urlprefix\url{https://link.aps.org/doi/10.1103/PhysRevLett.120.026801}.

\bibitem{PhysRevB.97.205136}
\bibinfo{author}{Khalaf, E.}
\newblock \bibinfo{title}{Higher-order topological insulators and
  superconductors protected by inversion symmetry}.
\newblock \textit{\bibinfo{journal}{Phys. Rev. B}}
  \textbf{\bibinfo{volume}{97}}, \bibinfo{pages}{205136}
  (\bibinfo{year}{2018}).
\newblock \urlprefix\url{https://link.aps.org/doi/10.1103/PhysRevB.97.205136}.

\bibitem{PhysRevB.98.201114}
\bibinfo{author}{Franca, S.}, \bibinfo{author}{van~den Brink, J.} \&
  \bibinfo{author}{Fulga, I.~C.}
\newblock \bibinfo{title}{An anomalous higher-order topological insulator}.
\newblock \textit{\bibinfo{journal}{Phys. Rev. B}}
  \textbf{\bibinfo{volume}{98}}, \bibinfo{pages}{201114}
  (\bibinfo{year}{2018}).
\newblock \urlprefix\url{https://link.aps.org/doi/10.1103/PhysRevB.98.201114}.

\bibitem{PhysRevB.97.205135}
\bibinfo{author}{Geier, M.}, \bibinfo{author}{Trifunovic, L.},
  \bibinfo{author}{Hoskam, M.} \& \bibinfo{author}{Brouwer, P.~W.}
\newblock \bibinfo{title}{Second-order topological insulators and
  superconductors with an order-two crystalline symmetry}.
\newblock \textit{\bibinfo{journal}{Phys. Rev. B}}
  \textbf{\bibinfo{volume}{97}}, \bibinfo{pages}{205135}
  (\bibinfo{year}{2018}).
\newblock \urlprefix\url{https://link.aps.org/doi/10.1103/PhysRevB.97.205135}.

\bibitem{PhysRevX.9.011012}
\bibinfo{author}{Trifunovic, L.} \& \bibinfo{author}{Brouwer, P.~W.}
\newblock \bibinfo{title}{Higher-order bulk-boundary correspondence for
  topological crystalline phases}.
\newblock \textit{\bibinfo{journal}{Phys. Rev. X}}
  \textbf{\bibinfo{volume}{9}}, \bibinfo{pages}{011012} (\bibinfo{year}{2019}).
\newblock \urlprefix\url{https://link.aps.org/doi/10.1103/PhysRevX.9.011012}.

\bibitem{Li_2DMater2021}
\bibinfo{author}{Li, T.}, \bibinfo{author}{Geier, M.}, \bibinfo{author}{Ingham,
  J.} \& \bibinfo{author}{Scammell, H.~D.}
\newblock \bibinfo{title}{Higher-order topological superconductivity from
  repulsive interactions in kagom{\'e} and honeycomb systems}.
\newblock \textit{\bibinfo{journal}{2D Mater.}} \textbf{\bibinfo{volume}{9}},
  \bibinfo{pages}{015031} (\bibinfo{year}{2021}).
\newblock \urlprefix\url{https://doi.org/10.1088/2053-1583/ac4060}.

\bibitem{Ghosh_PRB2021}
\bibinfo{author}{Ghosh, A.~K.}, \bibinfo{author}{Nag, T.} \&
  \bibinfo{author}{Saha, A.}
\newblock \bibinfo{title}{Hierarchy of higher-order topological superconductors
  in three dimensions}.
\newblock \textit{\bibinfo{journal}{Phys. Rev. B}}
  \textbf{\bibinfo{volume}{104}}, \bibinfo{pages}{134508}
  (\bibinfo{year}{2021}).
\newblock \urlprefix\url{https://link.aps.org/doi/10.1103/PhysRevB.104.134508}.

\bibitem{PhysRevResearch.2.012060}
\bibinfo{author}{Ahn, J.} \& \bibinfo{author}{Yang, B.-J.}
\newblock \bibinfo{title}{Higher-order topological superconductivity of
  spin-polarized fermions}.
\newblock \textit{\bibinfo{journal}{Phys. Rev. Research}}
  \textbf{\bibinfo{volume}{2}}, \bibinfo{pages}{012060} (\bibinfo{year}{2020}).
\newblock
  \urlprefix\url{https://link.aps.org/doi/10.1103/PhysRevResearch.2.012060}.

\bibitem{PhysRevResearch.2.043300}
\bibinfo{author}{Tiwari, A.}, \bibinfo{author}{Jahin, A.} \&
  \bibinfo{author}{Wang, Y.}
\newblock \bibinfo{title}{Chiral dirac superconductors: Second-order and
  boundary-obstructed topology}.
\newblock \textit{\bibinfo{journal}{Phys. Rev. Research}}
  \textbf{\bibinfo{volume}{2}}, \bibinfo{pages}{043300} (\bibinfo{year}{2020}).
\newblock
  \urlprefix\url{https://link.aps.org/doi/10.1103/PhysRevResearch.2.043300}.

\bibitem{Kitaev_2001}
\bibinfo{author}{Kitaev, A.~Y.}
\newblock \bibinfo{title}{Unpaired {M}ajorana fermions in quantum wires}.
\newblock \textit{\bibinfo{journal}{Phys.-Usp.}} \textbf{\bibinfo{volume}{44}},
  \bibinfo{pages}{131} (\bibinfo{year}{2001}).
\newblock \urlprefix\url{https://doi.org/10.1070/1063-7869/44/10s/s29}.

\bibitem{RevModPhys.80.1083}
\bibinfo{author}{Nayak, C.}, \bibinfo{author}{Simon, S.~H.},
  \bibinfo{author}{Stern, A.}, \bibinfo{author}{Freedman, M.} \&
  \bibinfo{author}{Das~Sarma, S.}
\newblock \bibinfo{title}{Non-{A}belian anyons and topological quantum
  computation}.
\newblock \textit{\bibinfo{journal}{Rev. Mod. Phys.}}
  \textbf{\bibinfo{volume}{80}}, \bibinfo{pages}{1083--1159}
  (\bibinfo{year}{2008}).
\newblock \urlprefix\url{https://link.aps.org/doi/10.1103/RevModPhys.80.1083}.

\bibitem{Sarma2015}
\bibinfo{author}{Sarma, S.~D.}, \bibinfo{author}{Freedman, M.} \&
  \bibinfo{author}{Nayak, C.}
\newblock \bibinfo{title}{Majorana zero modes and topological quantum
  computation}.
\newblock \textit{\bibinfo{journal}{npj Quantum Inf.}}
  \textbf{\bibinfo{volume}{1}}, \bibinfo{pages}{15001} (\bibinfo{year}{2015}).
\newblock \urlprefix\url{https://doi.org/10.1038/npjqi.2015.1}.

\bibitem{PhysRevX.6.031016}
\bibinfo{author}{Aasen, D.} \textit{et~al.}
\newblock \bibinfo{title}{Milestones toward {M}ajorana-based quantum
  computing}.
\newblock \textit{\bibinfo{journal}{Phys. Rev. X}}
  \textbf{\bibinfo{volume}{6}}, \bibinfo{pages}{031016} (\bibinfo{year}{2016}).
\newblock \urlprefix\url{https://link.aps.org/doi/10.1103/PhysRevX.6.031016}.

\bibitem{RevModPhys.87.137}
\bibinfo{author}{Elliott, S.~R.} \& \bibinfo{author}{Franz, M.}
\newblock \bibinfo{title}{Colloquium: {M}ajorana fermions in nuclear, particle,
  and solid-state physics}.
\newblock \textit{\bibinfo{journal}{Rev. Mod. Phys.}}
  \textbf{\bibinfo{volume}{87}}, \bibinfo{pages}{137} (\bibinfo{year}{2015}).
\newblock \urlprefix\url{https://link.aps.org/doi/10.1103/RevModPhys.87.137}.

\bibitem{Mohanta_EPL2014}
\bibinfo{author}{Mohanta, N.} \& \bibinfo{author}{Taraphder, A.}
\newblock \bibinfo{title}{Topological superconductivity and {M}ajorana bound
  states at the {LaAlO}$_3$/{SrTiO}$_3$ interface}.
\newblock \textit{\bibinfo{journal}{Europhys. Lett.}}
  \textbf{\bibinfo{volume}{108}}, \bibinfo{pages}{60001}
  (\bibinfo{year}{2014}).
\newblock \urlprefix\url{https://doi.org/10.1209/0295-5075/108/60001}.

\bibitem{PhysRevLett.105.077001}
\bibinfo{author}{Lutchyn, R.~M.}, \bibinfo{author}{Sau, J.~D.} \&
  \bibinfo{author}{Das~Sarma, S.}
\newblock \bibinfo{title}{Majorana fermions and a topological phase transition
  in semiconductor-superconductor heterostructures}.
\newblock \textit{\bibinfo{journal}{Phys. Rev. Lett.}}
  \textbf{\bibinfo{volume}{105}}, \bibinfo{pages}{077001}
  (\bibinfo{year}{2010}).
\newblock
  \urlprefix\url{https://link.aps.org/doi/10.1103/PhysRevLett.105.077001}.

\bibitem{Mourik_Science2012}
\bibinfo{author}{Mourik, V.} \textit{et~al.}
\newblock \bibinfo{title}{Signatures of {M}ajorana fermions in hybrid
  superconductor-semiconductor nanowire devices}.
\newblock \textit{\bibinfo{journal}{Science}} \textbf{\bibinfo{volume}{336}},
  \bibinfo{pages}{1003} (\bibinfo{year}{2012}).
\newblock
  \urlprefix\url{https://www.science.org/doi/abs/10.1126/science.1222360}.

\bibitem{Rokhinson2012}
\bibinfo{author}{Rokhinson, L.~P.}, \bibinfo{author}{Liu, X.} \&
  \bibinfo{author}{Furdyna, J.~K.}
\newblock \bibinfo{title}{The fractional a.c. josephson effect in a
  semiconductor--superconductor nanowire as a signature of {M}ajorana
  particles}.
\newblock \textit{\bibinfo{journal}{Nat. Phys.}} \textbf{\bibinfo{volume}{8}},
  \bibinfo{pages}{795} (\bibinfo{year}{2012}).
\newblock \urlprefix\url{https://doi.org/10.1038/nphys2429}.

\bibitem{Deng_Science2016}
\bibinfo{author}{Deng, M.~T.} \textit{et~al.}
\newblock \bibinfo{title}{Majorana bound state in a coupled quantum-dot
  hybrid-nanowire system}.
\newblock \textit{\bibinfo{journal}{Science}} \textbf{\bibinfo{volume}{354}},
  \bibinfo{pages}{1557} (\bibinfo{year}{2016}).
\newblock
  \urlprefix\url{https://www.science.org/doi/abs/10.1126/science.aaf3961}.

\bibitem{Desjardins2019}
\bibinfo{author}{Desjardins, M.~M.} \textit{et~al.}
\newblock \bibinfo{title}{Synthetic spin--orbit interaction for {M}ajorana
  devices}.
\newblock \textit{\bibinfo{journal}{Nat. Mater.}}
  \textbf{\bibinfo{volume}{18}}, \bibinfo{pages}{1060} (\bibinfo{year}{2019}).
\newblock \urlprefix\url{https://doi.org/10.1038/s41563-019-0457-6}.

\bibitem{PhysRevApplied.12.034048}
\bibinfo{author}{Mohanta, N.} \textit{et~al.}
\newblock \bibinfo{title}{Electrical control of {M}ajorana bound states using
  magnetic stripes}.
\newblock \textit{\bibinfo{journal}{Phys. Rev. Applied}}
  \textbf{\bibinfo{volume}{12}}, \bibinfo{pages}{034048}
  (\bibinfo{year}{2019}).
\newblock
  \urlprefix\url{https://link.aps.org/doi/10.1103/PhysRevApplied.12.034048}.

\bibitem{Mohanta_CommPhys2021}
\bibinfo{author}{Mohanta, N.}, \bibinfo{author}{Okamoto, S.} \&
  \bibinfo{author}{Dagotto, E.}
\newblock \bibinfo{title}{Skyrmion control of {M}ajorana states in planar
  {J}osephson junctions}.
\newblock \textit{\bibinfo{journal}{Comm. Phys.}} \textbf{\bibinfo{volume}{4}},
  \bibinfo{pages}{163} (\bibinfo{year}{2021}).
\newblock \urlprefix\url{https://doi.org/10.1038/s42005-021-00666-5}.

\bibitem{Herbrych2021}
\bibinfo{author}{Herbrych, J.}, \bibinfo{author}{{\'{S}}roda, M.},
  \bibinfo{author}{Alvarez, G.}, \bibinfo{author}{Mierzejewski, M.} \&
  \bibinfo{author}{Dagotto, E.}
\newblock \bibinfo{title}{Interaction-induced topological phase transition and
  {M}ajorana edge states in low-dimensional orbital-selective {M}ott
  insulators}.
\newblock \textit{\bibinfo{journal}{Nat. Commun.}}
  \textbf{\bibinfo{volume}{12}}, \bibinfo{pages}{2955} (\bibinfo{year}{2021}).
\newblock \urlprefix\url{https://doi.org/10.1038/s41467-021-23261-2}.

\bibitem{Song_PRL2017}
\bibinfo{author}{Song, Z.}, \bibinfo{author}{Fang, Z.} \&
  \bibinfo{author}{Fang, C.}
\newblock \bibinfo{title}{$(d\ensuremath{-}2)$-dimensional edge states of
  rotation symmetry protected topological states}.
\newblock \textit{\bibinfo{journal}{Phys. Rev. Lett.}}
  \textbf{\bibinfo{volume}{119}}, \bibinfo{pages}{246402}
  (\bibinfo{year}{2017}).
\newblock
  \urlprefix\url{https://link.aps.org/doi/10.1103/PhysRevLett.119.246402}.

\bibitem{PhysRevB.98.165144}
\bibinfo{author}{Wang, Y.}, \bibinfo{author}{Lin, M.} \&
  \bibinfo{author}{Hughes, T.~L.}
\newblock \bibinfo{title}{Weak-pairing higher order topological
  superconductors}.
\newblock \textit{\bibinfo{journal}{Phys. Rev. B}}
  \textbf{\bibinfo{volume}{98}}, \bibinfo{pages}{165144}
  (\bibinfo{year}{2018}).
\newblock \urlprefix\url{https://link.aps.org/doi/10.1103/PhysRevB.98.165144}.

\bibitem{PhysRevResearch.2.032068}
\bibinfo{author}{Pahomi, T.~E.}, \bibinfo{author}{Sigrist, M.} \&
  \bibinfo{author}{Soluyanov, A.~A.}
\newblock \bibinfo{title}{Braiding {M}ajorana corner modes in a second-order
  topological superconductor}.
\newblock \textit{\bibinfo{journal}{Phys. Rev. Research}}
  \textbf{\bibinfo{volume}{2}}, \bibinfo{pages}{032068} (\bibinfo{year}{2020}).
\newblock
  \urlprefix\url{https://link.aps.org/doi/10.1103/PhysRevResearch.2.032068}.

\bibitem{PhysRevResearch.2.043025}
\bibinfo{author}{Zhang, S.-B.} \textit{et~al.}
\newblock \bibinfo{title}{Topological and holonomic quantum computation based
  on second-order topological superconductors}.
\newblock \textit{\bibinfo{journal}{Phys. Rev. Research}}
  \textbf{\bibinfo{volume}{2}}, \bibinfo{pages}{043025} (\bibinfo{year}{2020}).
\newblock
  \urlprefix\url{https://link.aps.org/doi/10.1103/PhysRevResearch.2.043025}.

\bibitem{Kheirkhah_PRL2020}
\bibinfo{author}{Kheirkhah, M.}, \bibinfo{author}{Yan, Z.},
  \bibinfo{author}{Nagai, Y.} \& \bibinfo{author}{Marsiglio, F.}
\newblock \bibinfo{title}{First- and second-order topological superconductivity
  and temperature-driven topological phase transitions in the extended
  {H}ubbard model with spin-orbit coupling}.
\newblock \textit{\bibinfo{journal}{Phys. Rev. Lett.}}
  \textbf{\bibinfo{volume}{125}}, \bibinfo{pages}{017001}
  (\bibinfo{year}{2020}).
\newblock
  \urlprefix\url{https://link.aps.org/doi/10.1103/PhysRevLett.125.017001}.

\bibitem{PhysRevLett.121.096803}
\bibinfo{author}{Yan, Z.}, \bibinfo{author}{Song, F.} \& \bibinfo{author}{Wang,
  Z.}
\newblock \bibinfo{title}{Majorana corner modes in a high-temperature
  platform}.
\newblock \textit{\bibinfo{journal}{Phys. Rev. Lett.}}
  \textbf{\bibinfo{volume}{121}}, \bibinfo{pages}{096803}
  (\bibinfo{year}{2018}).
\newblock
  \urlprefix\url{https://link.aps.org/doi/10.1103/PhysRevLett.121.096803}.

\bibitem{PhysRevLett.121.186801}
\bibinfo{author}{Wang, Q.}, \bibinfo{author}{Liu, C.-C.}, \bibinfo{author}{Lu,
  Y.-M.} \& \bibinfo{author}{Zhang, F.}
\newblock \bibinfo{title}{High-temperature {M}ajorana corner states}.
\newblock \textit{\bibinfo{journal}{Phys. Rev. Lett.}}
  \textbf{\bibinfo{volume}{121}}, \bibinfo{pages}{186801}
  (\bibinfo{year}{2018}).
\newblock
  \urlprefix\url{https://link.aps.org/doi/10.1103/PhysRevLett.121.186801}.

\bibitem{PhysRevLett.122.126402}
\bibinfo{author}{Volpez, Y.}, \bibinfo{author}{Loss, D.} \&
  \bibinfo{author}{Klinovaja, J.}
\newblock \bibinfo{title}{Second-order topological superconductivity in
  $\ensuremath{\pi}$-junction {R}ashba layers}.
\newblock \textit{\bibinfo{journal}{Phys. Rev. Lett.}}
  \textbf{\bibinfo{volume}{122}}, \bibinfo{pages}{126402}
  (\bibinfo{year}{2019}).
\newblock
  \urlprefix\url{https://link.aps.org/doi/10.1103/PhysRevLett.122.126402}.

\bibitem{Cao2018}
\bibinfo{author}{Cao, Y.} \textit{et~al.}
\newblock \bibinfo{title}{Unconventional superconductivity in magic-angle
  graphene superlattices}.
\newblock \textit{\bibinfo{journal}{Nature}} \textbf{\bibinfo{volume}{556}},
  \bibinfo{pages}{43} (\bibinfo{year}{2018}).
\newblock \urlprefix\url{https://doi.org/10.1038/nature26160}.

\bibitem{PhysRevB.101.014501}
\bibinfo{author}{Sayyad, S.} \textit{et~al.}
\newblock \bibinfo{title}{Pairing and non-fermi liquid behavior in partially
  flat-band systems: Beyond nesting physics}.
\newblock \textit{\bibinfo{journal}{Phys. Rev. B}}
  \textbf{\bibinfo{volume}{101}}, \bibinfo{pages}{014501}
  (\bibinfo{year}{2020}).
\newblock \urlprefix\url{https://link.aps.org/doi/10.1103/PhysRevB.101.014501}.

\bibitem{PhysRevLett.126.027002}
\bibinfo{author}{Peri, V.}, \bibinfo{author}{Song, Z.-D.},
  \bibinfo{author}{Bernevig, B.~A.} \& \bibinfo{author}{Huber, S.~D.}
\newblock \bibinfo{title}{Fragile topology and flat-band superconductivity in
  the strong-coupling regime}.
\newblock \textit{\bibinfo{journal}{Phys. Rev. Lett.}}
  \textbf{\bibinfo{volume}{126}}, \bibinfo{pages}{027002}
  (\bibinfo{year}{2021}).
\newblock
  \urlprefix\url{https://link.aps.org/doi/10.1103/PhysRevLett.126.027002}.

\bibitem{Volovik2016}
\bibinfo{author}{{Heikkil{\"a}}, T.~T.} \& \bibinfo{author}{{Volovik}, G.~E.}
\newblock \bibinfo{title}{Flat bands as a route to high-temperature
  superconductivity in graphite}.
\newblock \textit{\bibinfo{journal}{Springer Series in Materials Science}}
  \textbf{\bibinfo{volume}{244}}, \bibinfo{pages}{123} (\bibinfo{year}{2016}).
\newblock \urlprefix\url{https://doi.org/10.1007/978-3-319-39355-1_6}.

\bibitem{Aoki2020}
\bibinfo{author}{Aoki, H.}
\newblock \bibinfo{title}{Theoretical possibilities for flat band
  superconductivity}.
\newblock \textit{\bibinfo{journal}{J. Supercond. Nov. Magn.}}
  \textbf{\bibinfo{volume}{33}}, \bibinfo{pages}{2341} (\bibinfo{year}{2020}).
\newblock \urlprefix\url{https://doi.org/10.1007/s10948-020-05474-6}.

\bibitem{PhysRevB.106.125155}
\bibinfo{author}{Mahyaeh, I.}, \bibinfo{author}{K\"ohler, T.},
  \bibinfo{author}{Black-Schaffer, A.~M.} \& \bibinfo{author}{Kantian, A.}
\newblock \bibinfo{title}{Superconducting pairing from repulsive interactions
  of fermions in a flat-band system}.
\newblock \textit{\bibinfo{journal}{Phys. Rev. B}}
  \textbf{\bibinfo{volume}{106}}, \bibinfo{pages}{125155}
  (\bibinfo{year}{2022}).
\newblock \urlprefix\url{https://link.aps.org/doi/10.1103/PhysRevB.106.125155}.

\bibitem{Peotta2015}
\bibinfo{author}{Peotta, S.} \& \bibinfo{author}{T{\"o}rm{\"a}, P.}
\newblock \bibinfo{title}{Superfluidity in topologically nontrivial flat
  bands}.
\newblock \textit{\bibinfo{journal}{Nat. Commun.}}
  \textbf{\bibinfo{volume}{6}}, \bibinfo{pages}{8944} (\bibinfo{year}{2015}).
\newblock \urlprefix\url{https://doi.org/10.1038/ncomms9944}.

\bibitem{PhysRevLett.124.167002}
\bibinfo{author}{Xie, F.}, \bibinfo{author}{Song, Z.}, \bibinfo{author}{Lian,
  B.} \& \bibinfo{author}{Bernevig, B.~A.}
\newblock \bibinfo{title}{Topology-bounded superfluid weight in twisted bilayer
  graphene}.
\newblock \textit{\bibinfo{journal}{Phys. Rev. Lett.}}
  \textbf{\bibinfo{volume}{124}}, \bibinfo{pages}{167002}
  (\bibinfo{year}{2020}).
\newblock
  \urlprefix\url{https://link.aps.org/doi/10.1103/PhysRevLett.124.167002}.

\bibitem{Sticlet_PRB2014}
\bibinfo{author}{Sticlet, D.}, \bibinfo{author}{Seabra, L.},
  \bibinfo{author}{Pollmann, F.} \& \bibinfo{author}{Cayssol, J.}
\newblock \bibinfo{title}{From fractionally charged solitons to {M}ajorana
  bound states in a one-dimensional interacting model}.
\newblock \textit{\bibinfo{journal}{Phys. Rev. B}}
  \textbf{\bibinfo{volume}{89}}, \bibinfo{pages}{115430}
  (\bibinfo{year}{2014}).
\newblock \urlprefix\url{https://link.aps.org/doi/10.1103/PhysRevB.89.115430}.

\bibitem{Verma_PNAS2021}
\bibinfo{author}{Verma, N.}, \bibinfo{author}{Hazra, T.} \&
  \bibinfo{author}{Randeria, M.}
\newblock \bibinfo{title}{Optical spectral weight, phase stiffness, and {T}$_c$
  bounds for trivial and topological flat band superconductors}.
\newblock \textit{\bibinfo{journal}{Proc. Natl. Acad. Sci. U.S.A.}}
  \textbf{\bibinfo{volume}{118}}, \bibinfo{pages}{e2106744118}
  (\bibinfo{year}{2021}).
\newblock \urlprefix\url{https://www.pnas.org/doi/abs/10.1073/pnas.2106744118}.

\bibitem{PhysRevB.84.241103}
\bibinfo{author}{Wang, F.} \& \bibinfo{author}{Ran, Y.}
\newblock \bibinfo{title}{Nearly flat band with chern number ${C}=2$ on the
  dice lattice}.
\newblock \textit{\bibinfo{journal}{Phys. Rev. B}}
  \textbf{\bibinfo{volume}{84}}, \bibinfo{pages}{241103}
  (\bibinfo{year}{2011}).
\newblock \urlprefix\url{https://link.aps.org/doi/10.1103/PhysRevB.84.241103}.

\bibitem{PhysRevB.102.045105}
\bibinfo{author}{Soni, R.}, \bibinfo{author}{Kaushal, N.},
  \bibinfo{author}{Okamoto, S.} \& \bibinfo{author}{Dagotto, E.}
\newblock \bibinfo{title}{Flat bands and ferrimagnetic order in electronically
  correlated dice-lattice ribbons}.
\newblock \textit{\bibinfo{journal}{Phys. Rev. B}}
  \textbf{\bibinfo{volume}{102}}, \bibinfo{pages}{045105}
  (\bibinfo{year}{2020}).
\newblock \urlprefix\url{https://link.aps.org/doi/10.1103/PhysRevB.102.045105}.

\bibitem{Gorbar_PRB2021}
\bibinfo{author}{Gorbar, E.~V.}, \bibinfo{author}{Gusynin, V.~P.} \&
  \bibinfo{author}{Oriekhov, D.~O.}
\newblock \bibinfo{title}{Gap generation and flat band catalysis in dice model
  with local interaction}.
\newblock \textit{\bibinfo{journal}{Phys. Rev. B}}
  \textbf{\bibinfo{volume}{103}}, \bibinfo{pages}{155155}
  (\bibinfo{year}{2021}).
\newblock \urlprefix\url{https://link.aps.org/doi/10.1103/PhysRevB.103.155155}.

\bibitem{PhysRevLett.12.474}
\bibinfo{author}{Schooley, J.~F.}, \bibinfo{author}{Hosler, W.~R.} \&
  \bibinfo{author}{Cohen, M.~L.}
\newblock \bibinfo{title}{Superconductivity in semiconducting
  {S}r{T}i{O}$_{3}$}.
\newblock \textit{\bibinfo{journal}{Phys. Rev. Lett.}}
  \textbf{\bibinfo{volume}{12}}, \bibinfo{pages}{474} (\bibinfo{year}{1964}).
\newblock \urlprefix\url{https://link.aps.org/doi/10.1103/PhysRevLett.12.474}.

\bibitem{PhysRev.163.380}
\bibinfo{author}{Koonce, C.~S.}, \bibinfo{author}{Cohen, M.~L.},
  \bibinfo{author}{Schooley, J.~F.}, \bibinfo{author}{Hosler, W.~R.} \&
  \bibinfo{author}{Pfeiffer, E.~R.}
\newblock \bibinfo{title}{Superconducting transition temperatures of
  semiconducting {S}r{T}i{O}$_{3}$}.
\newblock \textit{\bibinfo{journal}{Phys. Rev.}}
  \textbf{\bibinfo{volume}{163}}, \bibinfo{pages}{380} (\bibinfo{year}{1967}).
\newblock \urlprefix\url{https://link.aps.org/doi/10.1103/PhysRev.163.380}.

\bibitem{Horiguchi_1974}
\bibinfo{author}{Horiguchi, T.} \& \bibinfo{author}{Chen, C.~C.}
\newblock \bibinfo{title}{Lattice {G}reen's function for the diced lattice}.
\newblock \textit{\bibinfo{journal}{J. Math. Phys.}}
  \textbf{\bibinfo{volume}{15}}, \bibinfo{pages}{659} (\bibinfo{year}{1974}).
\newblock \urlprefix\url{https://doi.org/10.1063/1.1666703}.

\bibitem{Sutherland_PRB1986}
\bibinfo{author}{Sutherland, B.}
\newblock \bibinfo{title}{Localization of electronic wave functions due to
  local topology}.
\newblock \textit{\bibinfo{journal}{Phys. Rev. B}}
  \textbf{\bibinfo{volume}{34}}, \bibinfo{pages}{5208} (\bibinfo{year}{1986}).
\newblock \urlprefix\url{https://link.aps.org/doi/10.1103/PhysRevB.34.5208}.

\bibitem{Vidal_PRL1998}
\bibinfo{author}{Vidal, J.}, \bibinfo{author}{Mosseri, R.} \&
  \bibinfo{author}{Dou\ifmmode~\mbox{\c{c}}\else \c{c}\fi{}ot, B.}
\newblock \bibinfo{title}{Aharonov-{B}ohm cages in two-dimensional structures}.
\newblock \textit{\bibinfo{journal}{Phys. Rev. Lett.}}
  \textbf{\bibinfo{volume}{81}}, \bibinfo{pages}{5888} (\bibinfo{year}{1998}).
\newblock \urlprefix\url{https://link.aps.org/doi/10.1103/PhysRevLett.81.5888}.

\bibitem{Vidal_PRB2001}
\bibinfo{author}{Vidal, J.}, \bibinfo{author}{Butaud, P.},
  \bibinfo{author}{Dou\ifmmode~\mbox{\c{c}}\else \c{c}\fi{}ot, B.} \&
  \bibinfo{author}{Mosseri, R.}
\newblock \bibinfo{title}{Disorder and interactions in {A}haronov-{B}ohm
  cages}.
\newblock \textit{\bibinfo{journal}{Phys. Rev. B}}
  \textbf{\bibinfo{volume}{64}}, \bibinfo{pages}{155306}
  (\bibinfo{year}{2001}).
\newblock \urlprefix\url{https://link.aps.org/doi/10.1103/PhysRevB.64.155306}.

\bibitem{PhysRevB.94.125435}
\bibinfo{author}{Illes, E.} \& \bibinfo{author}{Nicol, E.~J.}
\newblock \bibinfo{title}{Magnetic properties of the
  $\ensuremath{\alpha}\ensuremath{-}{T}_{3}$ model: Magneto-optical
  conductivity and the {H}ofstadter butterfly}.
\newblock \textit{\bibinfo{journal}{Phys. Rev. B}}
  \textbf{\bibinfo{volume}{94}}, \bibinfo{pages}{125435}
  (\bibinfo{year}{2016}).
\newblock \urlprefix\url{https://link.aps.org/doi/10.1103/PhysRevB.94.125435}.

\bibitem{Raoux_PRL2014}
\bibinfo{author}{Raoux, A.}, \bibinfo{author}{Morigi, M.},
  \bibinfo{author}{Fuchs, J.-N.}, \bibinfo{author}{Pi\'echon, F.} \&
  \bibinfo{author}{Montambaux, G.}
\newblock \bibinfo{title}{From dia- to paramagnetic orbital susceptibility of
  massless fermions}.
\newblock \textit{\bibinfo{journal}{Phys. Rev. Lett.}}
  \textbf{\bibinfo{volume}{112}}, \bibinfo{pages}{026402}
  (\bibinfo{year}{2014}).
\newblock
  \urlprefix\url{https://link.aps.org/doi/10.1103/PhysRevLett.112.026402}.

\bibitem{PhysRevB.73.144511}
\bibinfo{author}{Rizzi, M.}, \bibinfo{author}{Cataudella, V.} \&
  \bibinfo{author}{Fazio, R.}
\newblock \bibinfo{title}{Phase diagram of the {B}ose-{H}ubbard model with
  $\tau_3$ symmetry}.
\newblock \textit{\bibinfo{journal}{Phys. Rev. B}}
  \textbf{\bibinfo{volume}{73}}, \bibinfo{pages}{144511}
  (\bibinfo{year}{2006}).
\newblock \urlprefix\url{https://link.aps.org/doi/10.1103/PhysRevB.73.144511}.

\bibitem{PhysRevLett.121.087001}
\bibinfo{author}{Xu, C.} \& \bibinfo{author}{Balents, L.}
\newblock \bibinfo{title}{Topological superconductivity in twisted multilayer
  {G}raphene}.
\newblock \textit{\bibinfo{journal}{Phys. Rev. Lett.}}
  \textbf{\bibinfo{volume}{121}}, \bibinfo{pages}{087001}
  (\bibinfo{year}{2018}).
\newblock
  \urlprefix\url{https://link.aps.org/doi/10.1103/PhysRevLett.121.087001}.

\bibitem{PhysRevLett.87.037004}
\bibinfo{author}{Gor'kov, L.~P.} \& \bibinfo{author}{Rashba, E.~I.}
\newblock \bibinfo{title}{Superconducting {2D} system with lifted spin
  degeneracy: {M}ixed singlet-triplet state}.
\newblock \textit{\bibinfo{journal}{Phys. Rev. Lett.}}
  \textbf{\bibinfo{volume}{87}}, \bibinfo{pages}{037004}
  (\bibinfo{year}{2001}).
\newblock
  \urlprefix\url{https://link.aps.org/doi/10.1103/PhysRevLett.87.037004}.

\bibitem{PhysRevB.97.214507}
\bibinfo{author}{Mohanta, N.}, \bibinfo{author}{Kampf, A.~P.} \&
  \bibinfo{author}{Kopp, T.}
\newblock \bibinfo{title}{Supercurrent as a probe for topological
  superconductivity in magnetic adatom chains}.
\newblock \textit{\bibinfo{journal}{Phys. Rev. B}}
  \textbf{\bibinfo{volume}{97}}, \bibinfo{pages}{214507}
  (\bibinfo{year}{2018}).
\newblock \urlprefix\url{https://link.aps.org/doi/10.1103/PhysRevB.97.214507}.

\bibitem{Li_CommPhys2023}
\bibinfo{author}{Li, S.}, \bibinfo{author}{Hu, L.-H.}, \bibinfo{author}{Zhang,
  R.-X.} \& \bibinfo{author}{Okamoto, S.}
\newblock \bibinfo{title}{Topological superconductivity from forward phonon
  scatterings}.
\newblock \textit{\bibinfo{journal}{Commun. Phys.}}
  \textbf{\bibinfo{volume}{6}}, \bibinfo{pages}{235} (\bibinfo{year}{2023}).
\newblock \urlprefix\url{https://doi.org/10.1038/s42005-023-01311-z}.

\bibitem{Vatsal_PRB2018}
\bibinfo{author}{Dwivedi, V.}, \bibinfo{author}{Hickey, C.},
  \bibinfo{author}{Eschmann, T.} \& \bibinfo{author}{Trebst, S.}
\newblock \bibinfo{title}{Majorana corner modes in a second-order {K}itaev spin
  liquid}.
\newblock \textit{\bibinfo{journal}{Phys. Rev. B}}
  \textbf{\bibinfo{volume}{98}}, \bibinfo{pages}{054432}
  (\bibinfo{year}{2018}).
\newblock \urlprefix\url{https://link.aps.org/doi/10.1103/PhysRevB.98.054432}.

\bibitem{Principi_PRB2021}
\bibinfo{author}{Wang, H.} \& \bibinfo{author}{Principi, A.}
\newblock \bibinfo{title}{Majorana edge and corner states in square and
  kagom{\'e} quantum spin-$\frac{3}{2}$ liquids}.
\newblock \textit{\bibinfo{journal}{Phys. Rev. B}}
  \textbf{\bibinfo{volume}{104}}, \bibinfo{pages}{214422}
  (\bibinfo{year}{2021}).
\newblock \urlprefix\url{https://link.aps.org/doi/10.1103/PhysRevB.104.214422}.

\bibitem{Kheirkhah_PRB2022}
\bibinfo{author}{Kheirkhah, M.}, \bibinfo{author}{Zhu, D.},
  \bibinfo{author}{Maciejko, J.} \& \bibinfo{author}{Yan, Z.}
\newblock \bibinfo{title}{Corner- and sublattice-sensitive {M}ajorana zero
  modes on the kagom{\'e} lattice}.
\newblock \textit{\bibinfo{journal}{Phys. Rev. B}}
  \textbf{\bibinfo{volume}{106}}, \bibinfo{pages}{085420}
  (\bibinfo{year}{2022}).
\newblock \urlprefix\url{https://link.aps.org/doi/10.1103/PhysRevB.106.085420}.

\bibitem{PhysRevLett.115.207002}
\bibinfo{author}{Kozii, V.} \& \bibinfo{author}{Fu, L.}
\newblock \bibinfo{title}{Odd-parity superconductivity in the vicinity of
  inversion symmetry breaking in spin-orbit-coupled systems}.
\newblock \textit{\bibinfo{journal}{Phys. Rev. Lett.}}
  \textbf{\bibinfo{volume}{115}}, \bibinfo{pages}{207002}
  (\bibinfo{year}{2015}).
\newblock
  \urlprefix\url{https://link.aps.org/doi/10.1103/PhysRevLett.115.207002}.

\bibitem{Hughes_Science2017}
\bibinfo{author}{Benalcazar, W.~A.}, \bibinfo{author}{Bernevig, B.~A.} \&
  \bibinfo{author}{Hughes, T.~L.}
\newblock \bibinfo{title}{Quantized electric multipole insulators}.
\newblock \textit{\bibinfo{journal}{Science}} \textbf{\bibinfo{volume}{357}},
  \bibinfo{pages}{61} (\bibinfo{year}{2017}).
\newblock
  \urlprefix\url{https://www.science.org/doi/abs/10.1126/science.aah6442}.

\bibitem{PhysRevB.82.184516}
\bibinfo{author}{Qi, X.-L.}, \bibinfo{author}{Hughes, T.~L.} \&
  \bibinfo{author}{Zhang, S.-C.}
\newblock \bibinfo{title}{Chiral topological superconductor from the quantum
  hall state}.
\newblock \textit{\bibinfo{journal}{Phys. Rev. B}}
  \textbf{\bibinfo{volume}{82}}, \bibinfo{pages}{184516}
  (\bibinfo{year}{2010}).
\newblock \urlprefix\url{https://link.aps.org/doi/10.1103/PhysRevB.82.184516}.

\bibitem{PhysRevResearch.2.023063}
\bibinfo{author}{Lesser, O.} \& \bibinfo{author}{Oreg, Y.}
\newblock \bibinfo{title}{Universal phase diagram of topological
  superconductors subjected to magnetic flux}.
\newblock \textit{\bibinfo{journal}{Phys. Rev. Research}}
  \textbf{\bibinfo{volume}{2}}, \bibinfo{pages}{023063} (\bibinfo{year}{2020}).
\newblock
  \urlprefix\url{https://link.aps.org/doi/10.1103/PhysRevResearch.2.023063}.

\bibitem{Clarke2013}
\bibinfo{author}{Clarke, D.~J.}, \bibinfo{author}{Alicea, J.} \&
  \bibinfo{author}{Shtengel, K.}
\newblock \bibinfo{title}{Exotic non-{A}belian anyons from conventional
  fractional quantum {H}all states}.
\newblock \textit{\bibinfo{journal}{Nat. Commun.}}
  \textbf{\bibinfo{volume}{4}}, \bibinfo{pages}{1348} (\bibinfo{year}{2013}).
\newblock \urlprefix\url{https://doi.org/10.1038/ncomms2340}.

\bibitem{PhysRevX.4.031009}
\bibinfo{author}{Vaezi, A.}
\newblock \bibinfo{title}{Superconducting analogue of the parafermion
  fractional quantum {H}all states}.
\newblock \textit{\bibinfo{journal}{Phys. Rev. X}}
  \textbf{\bibinfo{volume}{4}}, \bibinfo{pages}{031009} (\bibinfo{year}{2014}).
\newblock \urlprefix\url{https://link.aps.org/doi/10.1103/PhysRevX.4.031009}.

\bibitem{PhysRevResearch.2.013232}
\bibinfo{author}{Santos, L.~H.}
\newblock \bibinfo{title}{Parafermions in hierarchical fractional quantum
  {H}all states}.
\newblock \textit{\bibinfo{journal}{Phys. Rev. Research}}
  \textbf{\bibinfo{volume}{2}}, \bibinfo{pages}{013232} (\bibinfo{year}{2020}).
\newblock
  \urlprefix\url{https://link.aps.org/doi/10.1103/PhysRevResearch.2.013232}.

\bibitem{PhysRevB.104.235115}
\bibinfo{author}{Soni, R.} \textit{et~al.}
\newblock \bibinfo{title}{Multitude of topological phase transitions in
  bipartite dice and {L}ieb lattices with interacting electrons and {R}ashba
  coupling}.
\newblock \textit{\bibinfo{journal}{Phys. Rev. B}}
  \textbf{\bibinfo{volume}{104}}, \bibinfo{pages}{235115}
  (\bibinfo{year}{2021}).
\newblock \urlprefix\url{https://link.aps.org/doi/10.1103/PhysRevB.104.235115}.

\bibitem{Okamoto_CommPhys2022}
\bibinfo{author}{Okamoto, S.}, \bibinfo{author}{Mohanta, N.},
  \bibinfo{author}{Dagotto, E.} \& \bibinfo{author}{Sheng, D.~N.}
\newblock \bibinfo{title}{Topological flat bands in a kagom{\'e} lattice
  multiorbital system}.
\newblock \textit{\bibinfo{journal}{Commun. Phys.}}
  \textbf{\bibinfo{volume}{5}}, \bibinfo{pages}{198} (\bibinfo{year}{2022}).
\newblock \urlprefix\url{https://doi.org/10.1038/s42005-022-00969-1}.

\bibitem{PhysRevLett.125.247002}
\bibinfo{author}{Ortiz, B.~R.} \textit{et~al.}
\newblock \bibinfo{title}{{CsV}$_3${S}b$_5$: {A} {Z}$_2$ topological kagom{\'e}
  metal with a superconducting ground state}.
\newblock \textit{\bibinfo{journal}{Phys. Rev. Lett.}}
  \textbf{\bibinfo{volume}{125}}, \bibinfo{pages}{247002}
  (\bibinfo{year}{2020}).
\newblock
  \urlprefix\url{https://link.aps.org/doi/10.1103/PhysRevLett.125.247002}.

\bibitem{Zhao2021}
\bibinfo{author}{Zhao, H.} \textit{et~al.}
\newblock \bibinfo{title}{Cascade of correlated electron states in the
  kagom{\'e} superconductor {CsV}$_3${S}b$_5$}.
\newblock \textit{\bibinfo{journal}{Nature}} \textbf{\bibinfo{volume}{599}},
  \bibinfo{pages}{216--221} (\bibinfo{year}{2021}).
\newblock \urlprefix\url{https://doi.org/10.1038/s41586-021-03946-w}.

\bibitem{HU2022495}
\bibinfo{author}{Hu, Y.} \textit{et~al.}
\newblock \bibinfo{title}{Topological surface states and flat bands in the
  kagom{\'e} superconductor {CsV}$_3${S}b$_5$}.
\newblock \textit{\bibinfo{journal}{Sci. Bull.}} \textbf{\bibinfo{volume}{67}},
  \bibinfo{pages}{495--500} (\bibinfo{year}{2022}).
\newblock
  \urlprefix\url{https://www.sciencedirect.com/science/article/pii/S2095927321007349}.

\end{thebibliography}

\end{document}